\documentclass[a4paper,12pt]{article}
\usepackage{amssymb,amsmath,amsthm}
\usepackage{tikz}
\usepackage{hyperref} 

\numberwithin{equation}{section}

\newcommand{\RR}{{\mathbb{R}}}
\newcommand{\CC}{{\mathbb{C}}}
\newcommand{\ZZ}{{\mathbb{Z}}}

\newcommand{\pa}{\partial}
\newcommand{\ii}{{\rm i}}
\newcommand{\dd}{{\rm d}}
\newcommand{\sfrac}[2]{{\textstyle\frac{#1}{#2}}}
\newcommand{\Tr}{\mathrm{Tr}}

\newcommand{\half}{\frac{1}{2}}
\newcommand{\ep}{\varepsilon}
\newcommand{\quart}{\frac{1}{4}}
\theoremstyle{plain}

\theoremstyle{definition}

\title{
\vskip -100pt
\rightline{\small{DAMTP-2018-23}}
\vskip 50pt
\bf{Rolling Skyrmions and the Nuclear Spin-Orbit Force}}
\author{\bf{Derek Harland}$^\ast$
\ \bf{and Nicholas S. Manton}$^\dagger$
  \bigskip
  \\$^\ast$School of Mathematics, University of Leeds, UK
  \\email address: d.g.harland@leeds.ac.uk
  \bigskip
  \\$^\dagger$DAMTP, University of Cambridge, UK
  \\email address: N.S.Manton@damtp.cam.ac.uk
  }
\date{18th June 2018}

\begin{document}

\maketitle

\abstract{
We compute the nuclear spin-orbit coupling from the Skyrme model.
Previous attempts to do this were based on the product ansatz, and as
such were limited to a system of two well-separated nuclei.  Our
calculation utilises a new method, and is applicable to the
phenomenologically important situation of a single nucleon orbiting a
large nucleus.  We find that, to second order in perturbation theory,
the coefficient of the spin-orbit coupling induced by pion field
interactions has the wrong sign, but as the strength of the pion-nucleon
interactions increases the correct sign is recovered
non-perturbatively.
}

\section{Introduction}

The spin-orbit coupling is an important ingredient in nuclear
structure theory. Its presence implies that it is energetically
favourable for the spin and orbital angular momentum of a nucleon to
be aligned, particularly if this nucleon is moving close to the
surface of a larger nucleus. This explains the phenomenon of magic 
numbers, and it is important in the description of halo nuclei, to 
name just two examples. Unlike the spin-orbit force encountered in the study of
electron shells of an atom, the nuclear spin-orbit force is not merely
a relativistic effect but is caused by the strong interaction physics 
of nuclei.

The Skyrme model is an effective description of QCD, and a candidate
model of nuclei with a topologically conserved baryon number. 
It successfully accounts for phenomena such as the
stability of the alpha-particle, the long-range forces between nuclei,
and quantum numbers of excited states of very light nuclei.  Some of
the recent successes of the model include reproducing the excited
states of oxygen-16 \cite{hkm17} and carbon-12 \cite{rawlinson18}, nuclear binding energies 
of the correct magnitude \cite{ghs15}, accurately modelling neutron stars 
\cite{ans-gvw15} and a geometric explanation for certain magic nuclei \cite{manton17}.
  
However, one of the challenges in analysing the Skyrme model has been 
accounting for the spin-orbit coupling.  There have been several attempts to calculate 
the spin-orbit term in the nucleon-nucleon potential \cite{jjp85,nyman&riska88,riska&dannbom88,osyks88,asw93}. 
Most of these calculations were only valid for large separations and were also perturbative, and so corresponded to calculations taking into account one- and two-pion exchange. 
Almost all obtained the nucleon-nucleon spin-orbit coupling with the wrong sign, 
although \cite{jjp85} obtained the correct sign by introducing additional mesons in the model.

The conventional description of the spin-orbit force is in the framework of relativistic mean field theory \cite{reinhard89},
which couples nucleons to several mesons (including the pion, $\sigma$, $\rho$ and $\omega$). 
An interesting perspective was put forward by Kaiser and Weise \cite{kaiser&weise08}:
they argued that the spin-orbit coupling receives several contributions, including a wrong-sign contribution from pion exchange; this 
is compensated by other effects, including meson exchange and three-body forces.  This seems to be related to the sign problem in the Skyrme model.

In this article we investigate in a novel way how a short-range 
spin-orbit coupling arises in the Skyrme model. Unlike relativistic mean field theory, 
our calculation is non-relativistic and incorporates pions but no other mesons. 
Our calculations are for a somewhat simplified model,
but we hope this model captures the essence of the effect.
Our main discovery is that the sign of the spin-orbit
coupling is wrong at weak coupling, where a perturbative approach
would be valid. However, the sign is correct when the coupling
between the nucleon and the surface of the nucleus with which it
interacts is stronger.

A key property of a Skyrmion, distinguishing it from an elementary
nucleon, is that it has orientational degrees of freedom. It is a
spherical rigid rotor. After quantisation \cite{anw83}, the basic states are
nucleons with spin $\half$, but there are also excited states with
spin $\frac{3}{2}$ corresponding to Delta resonances, and further
states of higher spin and higher energy that play no
significant role. The states simultaneously have isospin 
quantum numbers (isospin $\half$ for the nucleons and $\frac{3}{2}$ 
for the Deltas). In our model, a dynamical Skyrmion interacts
quantum mechanically with a background multi-Skyrmion field modelling the
nuclear surface. The interaction involves a potential that depends on the
Skyrmion orientation and its position, and the potential has a strength
parameter that we consider as adjustable. When the parameter is small, a
perturbative treatment works. However, the spin-orbit coupling has the 
wrong sign in this regime. When the parameter is larger (but not too 
large), the spin-orbit coupling for the Skyrmion has the correct sign.

Indeed, in this latter regime, a better approximation to 
the Skyrmion wavefunction is to say that the orientation has its 
probability concentrated near the minimum of the orientational potential, 
with this minimum varying with the Skyrmion's location on the surface. 
The quantum state is now close to the classical picture of a Skyrmion 
rolling over the nuclear surface, maintaining a minimal orientational 
potential energy. This classical rolling motion gives the correct sign 
for the spin-orbit coupling. In earlier work, Halcrow and one of 
the present authors investigated a model of this type 
\cite{halcrow&manton15}, but 
they only treated the case of a disc interacting with another disc in
two dimensions. When the potential is strong, the model becomes a quantised 
version of cog wheels rolling around each other. Here we do better, 
by treating a realistic three-dimensional Skyrmion interacting quantum
mechanically with a nuclear surface. However, we still need to make 
various approximations. For example, we assume the height of the 
Skyrmion above the surface is fixed.

Our analysis is based on the following well-known interpretation of the
phenomenological spin-orbit coupling. Consider a nucleon near the surface of a
spherical nucleus. Suppose that in addition to the usual kinetic
terms, the hamiltonian for the nucleon contains a term of the form 
\begin{equation}
\label{coupling term}
a\,\vec{S}.\vec{N}\times\vec{P} \,,
\end{equation}
where $a$ is a parameter, $\vec{S}$ is the spin of the nucleon,
$\vec{P}$ is its momentum, and $\vec{N}$ is an inward-pointing vector 
normal to the surface, which may be interpreted as the gradient of the 
density of nuclear matter.  Since the position vector $\vec{r}$ of 
the nucleon equals $-r\vec{N}/|\vec{N}|$, this term equals
$-(a|\vec{N}|/r)\vec{S}.\vec{L}$, where $\vec{L}=\vec{r}\times\vec{P}$
is the orbital angular momentum of the nucleon.  This is the usual form of the
spin-orbit coupling.  In order to give the correct magic numbers,
the spin-orbit coupling must prefer spin and angular momentum to be
aligned rather than anti-aligned, so the parameter $a$ needs to be
positive. The advantage for us of the formula (\ref{coupling term}) is
that it applies when the nucleon is interacting with an essentially 
flat nuclear surface, as in the model we will discuss below. We will
refer to (\ref{coupling term}) as the spin-momentum coupling. Note 
that $\vec{N}$ is essential here, and implies that there is no
coupling for an isolated nucleon, nor for a nucleon deeply embedded 
inside a nucleus.

There are two practical difficulties with this approach: the first 
is that the interaction between Skyrmions and multi-Skyrmions is
poorly understood at short distances, and the second is that the 
complicated spatial structure of known multi-Skyrmions with finite
baryon number would make the calculations laborious.  We solve the
first of these problems by working in the lightly bound version of the
Skyrme model \cite{ghkms17}, for which multi-Skyrmions and their 
interactions are accurately captured by a point particle 
description, although the particles still
have orientational degrees of freedom.  We solve the
second problem by supposing that the multi-Skyrmion representing the
core of the nucleus is large, and approximating its surface by a plane.
Since Skyrmions in the lightly bound model naturally arrange
themselves to sit at vertices of an FCC lattice, this surface has a 
high degree of symmetry, making the calculation tractable.

In the next section we review the 2D toy model of 
\cite{halcrow&manton15}, but in a modified
and simplified form. Here the dynamical, Skyrmion-like object is a coloured 
disc, and it moves in the background of a straight, periodically coloured rail,
rather than around a larger coloured disc as in 
\cite{halcrow&manton15}. The potential
depends on the colour difference between the disc and the rail at
their closest points. The translational and rotational motion of the
disc is quantised, and we compare the result of a perturbative treatment, valid
when the potential is weak but which leads to a spin-momentum coupling
of the wrong sign, with a non-perturbative approach that can deal 
with stronger coupling but is still algebraically straightforward. 
The price to pay for working non-perturbatively is that we must 
assume that the moment of inertia of the disc is small; in our perturbative
calculation, no such assumption is necessary. The strong coupling result
gives the correct sign for the spin-momentum coupling. 
In the later sections we perform similar calculations in the more
realistic 3D setting. Here, the Skyrmion is visualised as a
coloured sphere moving relative to a coloured surface, and the
potential again depends on the colour difference at the closest
points. The calculations can be done by
hand, exploiting the assumed lattice symmetries of the (planar)
nuclear surface, but are nevertheless considerably more complicated. 
The reader may wish to skip the details here. 

\section{Disc on a rail}

\begin{figure}
\begin{center}
\includegraphics[width=6in]{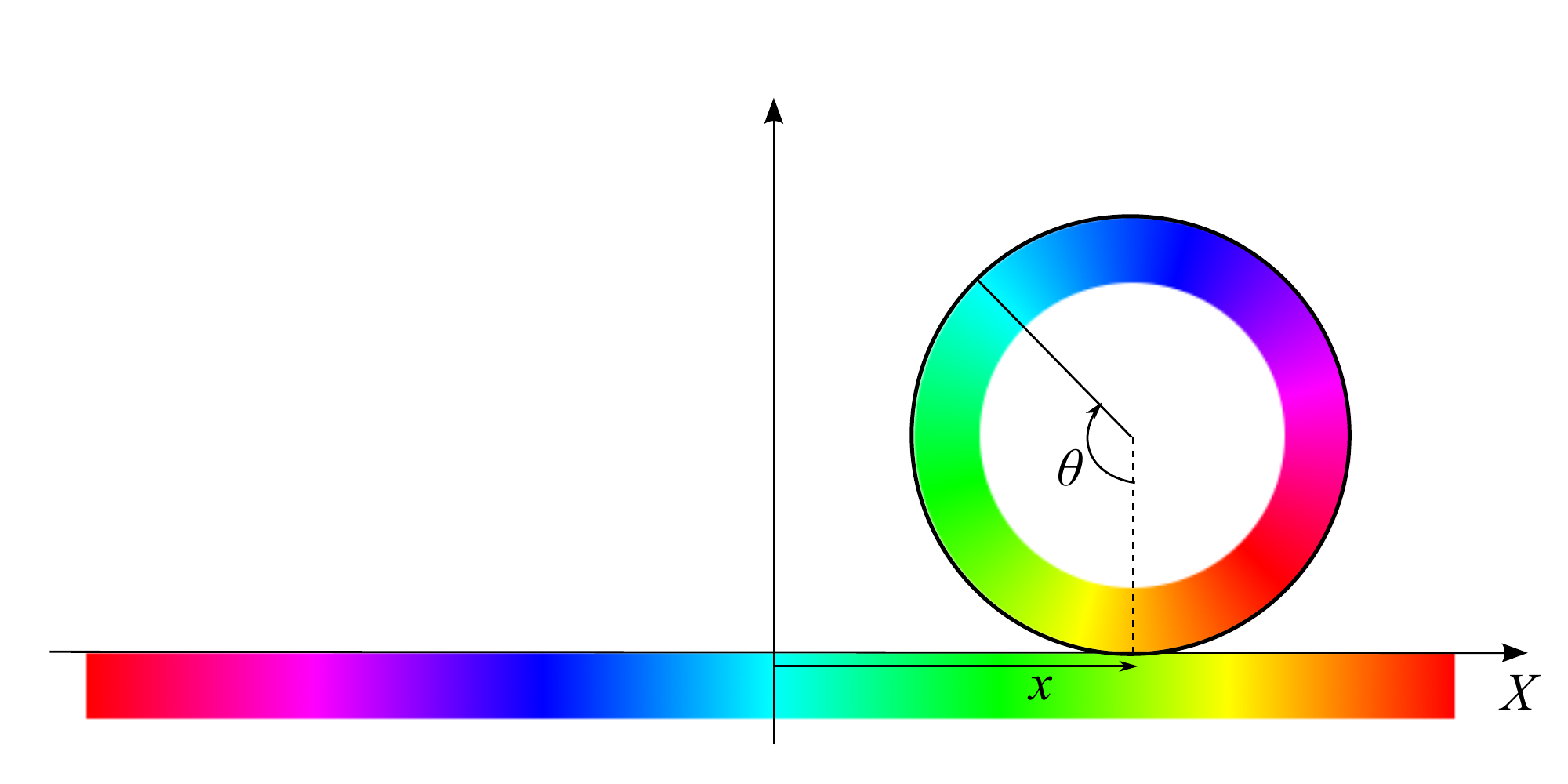}
\end{center}
\caption{Coloured disc on a fixed coloured rail. One period of the
  rail colouring is shown.}
\end{figure}

We start with a two-dimensional toy model of spin-momentum coupling, rather
similar to what was analysed in \cite{halcrow&manton15}. 
Consider a vertical disc
at a fixed height above a fixed, straight rail. The disc can move
along the rail and also rotate. Both the edge of the disc and the rail are 
coloured, and the potential energy is a periodic function of the colour
difference at their closest points. When the colours match, the 
potential energy is lowest. Let us assume that the disc is coloured so 
that for the potential to remain at its lowest value as the disc moves
classically, the disc needs to roll along the rail. See Figure 1. This model
is similar to a cog on a rack rail, which can only roll, but not slip.
Classically there is spin-momentum coupling, as the (clockwise) spin
of a rolling cog is a positive multiple of its linear momentum.

Let $X$ be a linear coordinate along the rail. The colour $\chi$ 
along the rail is an angular field variable, and as with an ordinary 
angle we assume $\chi$ takes any real value and identify values that 
differ by $2\pi$. We suppose that $\chi = X$, so the colour is
periodic along the rail, with period $2\pi$. Let the disc have radius 
1 and assume that when it is in its standard orientation, the colour
is the same as the angle around the disc measured from the bottom in an 
anticlockwise direction, i.e. the colour is $\chi$ at angle $\chi$.

Suppose now that the position and orientation of the disc are
$(x,\theta)$, where $x$ is the location of the centre of the disc,
projected down to the $X$-axis, and $\theta$ is the angle by which the disc
is rotated clockwise relative to its standard orientation. The bottom of
the disc then has colour $\theta$, and the rail under this point has colour
$x$. We suppose the potential energy of the disc in this configuration is 
$-V_0 \cos(x - \theta)$ with $V_0 \ge 0$.

We next introduce some dynamics. Suppose the disc has unit mass, and 
moment of inertia $\Lambda$, so the Lagrangian for its motion is
\begin{equation}
L = \half {\dot x}^2 + \half \Lambda {\dot\theta}^2 + V_0 \cos(x - \theta) \,.
\end{equation}
The equations of motion are
\begin{equation}
\ddot x = -V_0 \sin(x - \theta) \,, \quad 
\Lambda \ddot \theta = V_0 \sin(x - \theta) \,.
\end{equation}
Note that as the potential only depends on $x - \theta$, there is a
conserved quantity $\dot x + \Lambda \dot\theta$. One solution of the
equations is $x = \mu t$, $\theta = \mu t$ for any constant 
$\mu$ -- this is rolling motion. 
      
The conjugate momenta to $x$ and $\theta$ are
\begin{equation}
p = \dot x \,, \quad s = \Lambda \dot \theta \,,
\end{equation}
and the Hamiltonian is 
\begin{equation}
H = \half p^2 + \frac{1}{2\Lambda} s^2 - V_0 \cos(x - \theta) \,,
\label{Ham}
\end{equation}
with conserved quantity $p + s$.

We now quantise. Stationary wavefunctions are of the form 
$\Psi(x,\theta)$, and the momentum and spin operators are
\begin{equation}
p = -i\frac{\partial}{\partial x} \,, \quad 
s = -i\frac{\partial}{\partial\theta} \,.
\end{equation}
The stationary Schr\"odinger equation is
\begin{equation}
\left( -\half\frac{\partial^2}{\partial x^2}
-\frac{1}{2\Lambda} \frac{\partial^2}{\partial \theta^2}
- V_0 \cos(x - \theta) \right)\Psi = E \Psi \,,
\label{Schr}
\end{equation}
where the operator on the left hand side is the Hamiltonian (\ref{Ham})
expressed in terms of the momentum and spin operators.

The configuration space of the disc has first homotopy group $\ZZ$, so
wavefunctions can acquire a phase when $\theta \to \theta +
2\pi$. Bearing in mind that we are modelling a fermionic nucleon
interacting with a large nucleus, we choose this phase to be
$\pi$. Wavefunctions then have a Fourier expansion
\begin{equation}
\Psi(x,\theta) = \sum_{n \ \rm{odd}} \psi_n(x) e^{i\half n\theta} \,,
\end{equation}
a superposition of half-integer spin states. 

The free motion, in the absence of the potential, has separately
conserved momentum $p$ and spin $s$, and the basic stationary state is 
\begin{equation}
\Psi(x,\theta) = e^{ipx}e^{is\theta} \,,
\end{equation}
where $p$ is arbitrary and $s$ is half-integer. This state has energy
\begin{equation}
E = \half p^2 + \frac{1}{2\Lambda} s^2 \,.
\end{equation} 
We now suppose that $\Lambda$ is small, so that $\frac{1}{\Lambda}$ is large
compared to $V_0$ and to $p^2$. The expressions we derive later will 
only be valid provided $p^2 \ll \frac{1}{\Lambda}$. In this regime, the
low energy states are those with $s = \pm \half$. This is physically
what we are interested in. Spin $\frac{3}{2}$ nucleons (i.e. Delta
resonances) have energy about 300 MeV greater than spin $\half$
nucleons, and spin-orbit energies are much less than this, of
order 1 MeV. So we mostly neglect the small parts of the wavefunction with
$s = \pm \frac{3}{2}$ or larger.

Because of the restriction to $n = \pm 1$ states, i.e. those with
$s = \pm \half$, the wavefunction reduces to
\begin{equation}
\Psi(x,\theta) = \psi_1(x)e^{i\half\theta} +
\psi_{-1}(x)e^{-i\half\theta} \,.
\label{wavefncoupled}
\end{equation}
A stationary state like this is not strictly compatible with the 
Schr\"odinger equation, because the potential couples it to
$s = \pm \frac{3}{2}$ states. We can deal with this by calculating the matrix
form of the Hamiltonian restricted to this subspace of
wavefunctions. Recall that there is the conserved
quantity $p + s$. This implies that if $\psi_1(x) = e^{ipx}$ then
$\psi_{-1}(x) = A e^{ip'x}$, where $p' = p + 1$, for some 
amplitude $A$. Momentum $p$ itself is not a good label for states,
but instead we can use $r = p + s$, where $r$ takes any value in 
the range $(-\infty, \infty)$. The wavefunction (\ref{wavefncoupled}) 
becomes, for a definite value of $r$, 
\begin{equation}
\Psi(x,\theta) = e^{i(r-\half)x}e^{i\half\theta} 
+ Ae^{i(r+\half)x}e^{-i\half\theta} \,.
\label{wavefnr}
\end{equation}
Alternatively, the crystal momentum $k$ could be defined to 
be $p$ mod 1 and the (first) Brillouin zone to be $-\half \le k \le
\half$, but because of the restricted range of spins, we do not need the 
formalism of Bloch states mixing momentum $p$ with all 
its integer shifts.

We now work with basis states $\frac{1}{2\pi} e^{i(r-\half)x}e^{i\half\theta}$
and $\frac{1}{2\pi} e^{i(r+\half)x}e^{-i\half\theta}$. These are
normalised in $ \{ 0 \le x \le 2\pi \,, 0 \le \theta \le 2\pi \}$. The
matrix elements of the Hamiltonian (\ref{Ham}), or equivalently the
operator on the left of (\ref{Schr}), are
\begin{equation}
H_{2\times 2} = \begin{pmatrix} \half (r-\half)^2 + \frac{1}{8\Lambda} & -\half V_0 \\
-\half V_0 & \half (r+\half)^2 + \frac{1}{8\Lambda} \\ \end{pmatrix} \,,
\end{equation}
where the diagonal terms are kinetic contributions. The upper off-diagonal 
term comes from the matrix element of the potential
\begin{equation}
\frac{1}{(2\pi)^2} \int_0^{2\pi}\int_0^{2\pi}
e^{-i(r-\half)x}e^{-i\half\theta} (- V_0 \cos(x - \theta))
e^{i(r+\half)x}e^{-i\half\theta} \, dx d\theta  = -\half V_0 \,,
\label{Ham2*2}
\end{equation}
and the lower off-diagonal term is the same, by hermiticity. The potential 
makes no contribution to the diagonal terms.

It is now convenient to express the energy eigenvalues $E$ of $H_{2\times 2}$
as $E = \half\ep + \frac{1}{8\Lambda}$. The matrix with eigenvalues $\ep$ is 
\begin{equation}
\widetilde{H}_{2\times 2} = \begin{pmatrix} (r - \half)^2 & -V_0 \\
-V_0 & (r + \half)^2 \\ \end{pmatrix} \,,
\label{Hamred}
\end{equation}
and the eigenvalue equation $\det(\widetilde{H}_{2\times 2} - \ep {\bf 1}) = 0$
reduces to
\begin{equation} 
\ep^2 - 2\ep\left(r^2 + \quart\right) + \left(r^2 - \quart\right)^2 
- V_0^2 = 0 \,,
\end{equation}
with solutions
\begin{equation} 
\ep_{\pm}(r) = r^2 + \quart \pm \sqrt{r^2 + V_0^2} \,.
\label{evalues}
\end{equation}
The spectrum has two branches, the lower branch $\ep_-(r)$ 
and the upper branch $\ep_+(r)$, and is symmetric under $r \to -r$. 
When $V_0 = 0$ the spectrum simplifies to $\ep(r) = (r \pm \half)^2$, 
whose graph consists of two intersecting parabolas, with minima at 
$r = -\half$ and $r = \half$, and a crossover at $r=0$.

We are mainly interested in low energy states on the lower branch,
near the minima of $\ep_-(r)$.
There is an important bifurcation at a critical strength of the
potential, $V_0 = \half$. For $V_0 < \half$, $\ep_-$ has two minima at
$r = \pm \sqrt{\quart - V_0^2}$ and a local maximum at $r=0$. For 
$V_0 > \half$, there is just one minimum at $r=0$; here $p = \pm \half$,
so the crystal momentum $k$ is located on the boundary of the Brillouin
zone. The upper branch $\ep_+(r)$ has simpler behaviour, as it just 
has a minimum at $r=0$ for all positive $V_0$. Figure 2 shows graphs 
of the eigenvalue spectrum for two typical values of $V_0$.

\begin{figure}[htb]
\begin{center}
\includegraphics{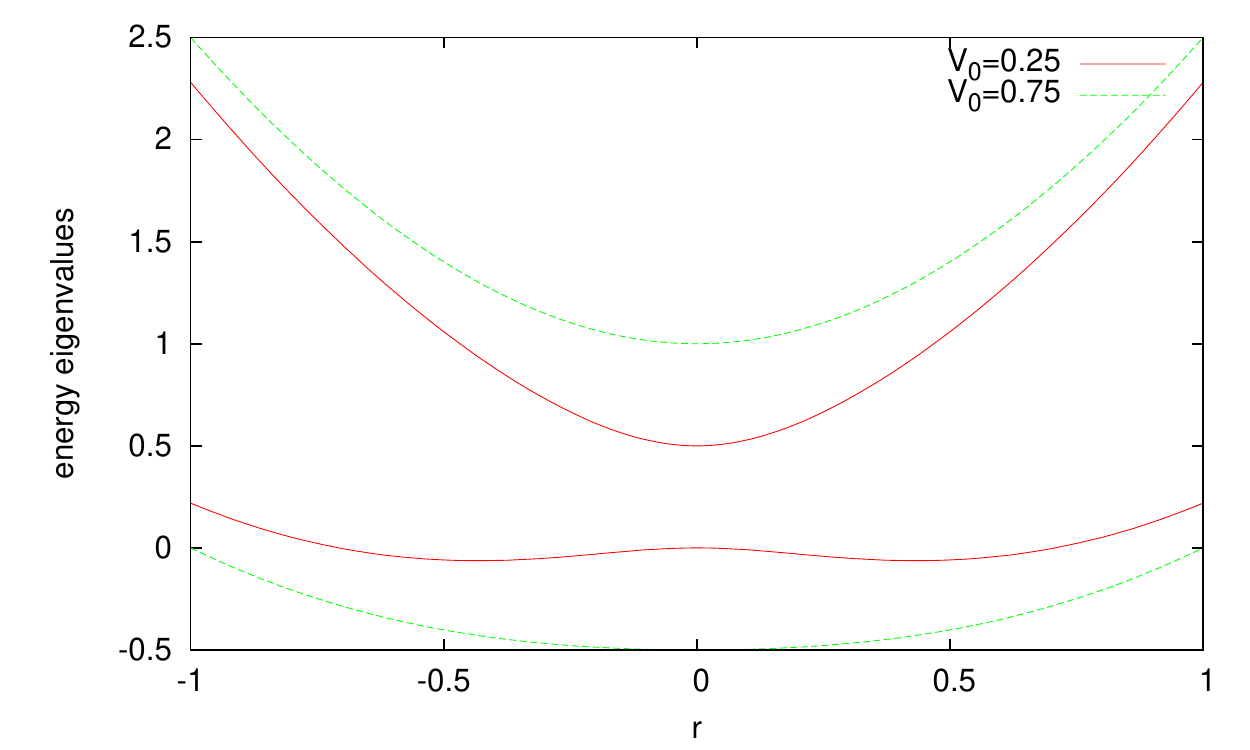}
\end{center}
\caption{Eigenvalues $\ep_-(r)$ (lower curves) and $\ep_+(r)$
(upper curves) of the matrix $\widetilde{H}_{2\times 2}$ for two
values of the parameter $V_0$.}
\end{figure}

Recall the form of the (unnormalised) wavefunction
(\ref{wavefnr}). We evaluate $A$ using the condition that
$1 \choose A$ is the eigenvector of the matrix (\ref{Hamred}). On the
lower branch of the spectrum
\begin{equation}
A = \frac{1}{V_0}\left(\sqrt{r^2 + V_0^2} - r \right) \,,  
\label{Alower}
\end{equation}
and on the upper branch
\begin{equation}
A = -\frac{1}{V_0}\left(\sqrt{r^2 + V_0^2} + r \right) \,.  
\label{Aupper}
\end{equation}
Note that $|A|^2 = 1$ at $r=0$ on both branches, so spins $\pm \half$
are superposed there with equal probability. On the lower branch, the
total (unnormalised) wavefunction at $r=0$ is $2\cos \half(x-\theta)$, 
so the highest probability occurs for $\theta = x$, where the disc is 
oriented so as to minimise the potential energy. This is compatible 
with a rolling motion.

Except in cases where $|A|$ is very close to 0, or much larger than 1,
the quantum states of the disc cannot be thought of as having a
definite momentum $p$ or spin $s$, because the potential strongly
superposes states where these have different values. So to consider 
the correlation between the momentum and spin, we work with their 
expectation values $\langle P \rangle$ and $\langle S \rangle$.

The expectation value of the spin is
\begin{equation}
\langle S \rangle = \frac{\half + |A|^2(-\half)}{1 + |A|^2}
\end{equation} 
where $A$ is given by expressions (\ref{Alower}) and (\ref{Aupper}), 
respectively, on the lower and upper branches. The expectation value of
momentum follows immediately, as $p+s = r$ for both contributing
states in (\ref{wavefnr}), so
\begin{equation}
\langle P \rangle = r - \langle S \rangle \,.
\end{equation}
Graphs of $\langle P \rangle$ and $\langle S \rangle$ as functions of
$r$ are shown in Figure 3. They are plotted together with $\ep_-$ for 
states on the lower branch, for the typical values of $V_0$ we selected 
before; for states on the upper branch, they are plotted together with
$\ep_+$.

\begin{figure}[htb]
\begin{center}
\includegraphics{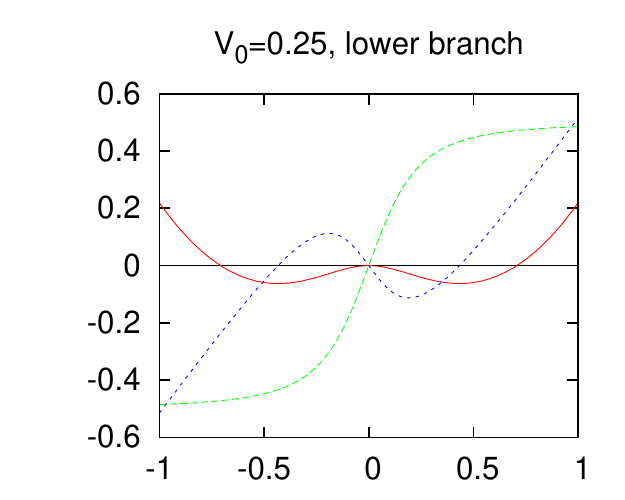}
\includegraphics{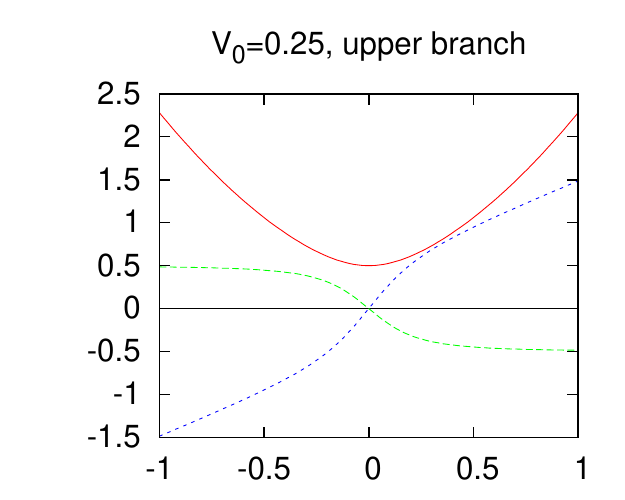}
\includegraphics{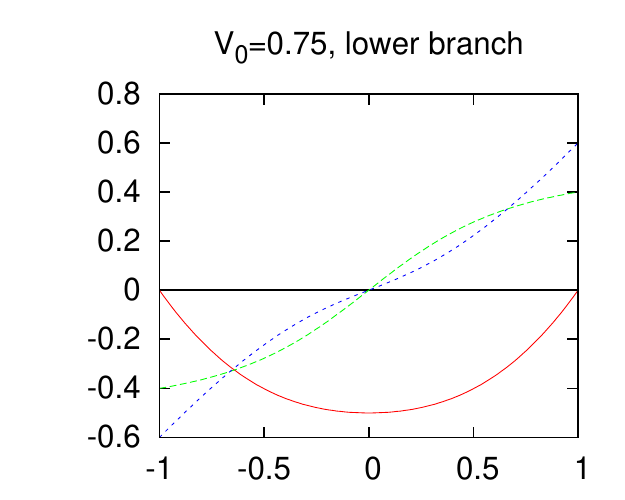}
\includegraphics{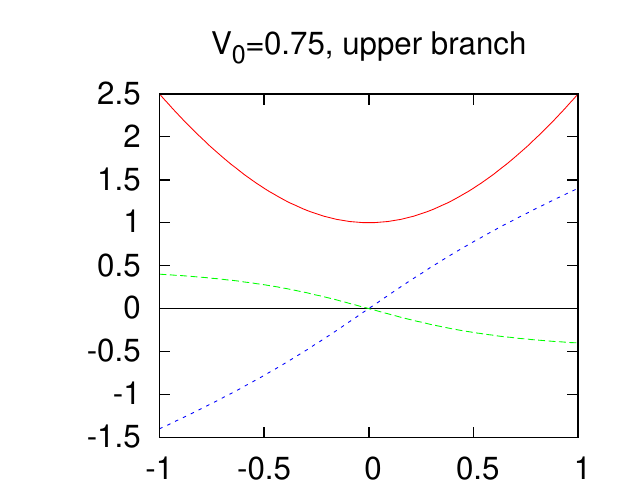}
\end{center}
\caption{Expectation values of the scaled energy $\ep$ (solid red), 
spin $\langle S \rangle$ (long-dashed green), and momentum $\langle P
\rangle$ (short-dashed blue) as functions of $r$, for the eigenstates of 
$\widetilde{H}_{2\times 2}$.}
\end{figure}

Interesting to note is that $\langle P \rangle$ vanishes 
wherever $E$ (or equivalently $\ep$) is stationary with respect to 
$r$, as one can see from the graphs. This is because
\begin{equation} 
\frac{d}{dr} {H_{2\times 2}} = \begin{pmatrix} r - \half & 0 \\
0 & r + \half \\ \end{pmatrix} \,,
\label{diffHamred}
\end{equation}
and the right hand side is the matrix form of the momentum operator. Taking
expectation values gives the result. (One also needs to use the identity
$\langle \frac{d}{dr}\Psi | \Psi \rangle 
+ \langle \Psi | \frac{d}{dr}\Psi \rangle = 0$ for normalised states.) 

When the potential is relatively weak, such that $V_0 < \half$, 
then $\langle P \rangle$ passes through 0 at the non-zero
minima of $\ep_-$ on the lower energy branch. On the other hand 
$\langle S \rangle$ does not change sign near here. The signs of 
$\langle P \rangle$ and $\langle S \rangle$ are therefore not strongly 
correlated for these low energy states, and we conclude that
for weak potentials there is no significant spin-momentum coupling.
Near $r=0$, where $\ep_-$ has a local maximum, $\langle P \rangle$ 
and $\langle S \rangle$ have opposite signs, so momentum and spin are
anticorrelated. This is the opposite of the classical correlation of
momentum and spin for a rolling motion. Similarly, on the upper energy
branch, the expectations of momentum and spin have opposite signs for
all $r$, so they are anticorrelated.

When the potential is stronger, such that $V_0 > \half$, we find 
the correlation we are seeking. Here, the
low energy states on the lower branch are near $r=0$, and we see
that $\langle P \rangle$ and $\langle S \rangle$ have the same
sign. It is straightforward to estimate these quantities analytically
for small $r$. They are
\begin{equation}
\label{lineariseExpectations}
\langle S \rangle \simeq \frac{1}{2V_0} \, r \quad {\rm and} \quad
\langle P \rangle \simeq \left(1 - \frac{1}{2V_0}\right)r \,,
\end{equation}
and for $V_0 > \half$ both their slopes with respect to $r$ are
positive. In fact, because $\langle P \rangle$ is zero only at $r=0$ 
when $V_0 > \half$, there is a spin-momentum correlation of the desired 
sign for all $r$, on the lower branch. On the other hand, the momentum 
and spin are anticorrelated for all $r$ on the upper branch.

The conclusion is that the potential has to be quite strong to achieve
the spin-momentum coupling for quantum states that mimics the classical 
phenomenon of rolling motion for a cog. As in the usual model of
spin-orbit coupling for a spin $\half$ particle, there are two states, a
lower energy state with a positive correlation, and a higher energy
state with an anticorrelation.

\subsection{Perturbation theory}

We have just seen that the spin-momentum correlation has the desired
form only when the potential is quite strong. Nevertheless, it is of
some interest to calculate what happens in perturbation theory. When
the potential is weak, we can calculate the energy spectrum to second
order in perturbation theory, treating $V_0$ as small. The perturbative result 
overlaps what we have already calculated, and we can allow for the
possibility that the moment of inertia $\Lambda$ is not small. This is a
useful check on our calculations, both for the two-dimensional disc, and
later, when we consider three-dimensional Skyrmion dynamics.

When $V_0 = 0$, the eigenstates of the Hamiltonian are 
$\Psi_0(x,\theta) = e^{ipx}e^{is\theta}$, with definite momentum and
spin, and energy $E = \half p^2 + \frac{1}{2\Lambda} s^2$. Low energy 
states are those with $s = \pm \half$ and $p \simeq 0$. These are near
the centre of the Brillouin zone. Let us focus on 
the states with $s = \half$ (the results are similar for $s = -\half$), whose
energy is $\half p^2 + \frac{1}{8\Lambda}$. Recall that when the potential 
is included, there is still the good quantum number $r = p + s$, so the 
states that we are focussing on have $r = p + \half \simeq \half$.

The effect of the cosine potential $-V_0\cos(x - \theta)$, at leading
order, is to mix the unperturbed state $\Psi_0 = e^{ipx}e^{i\half\theta}$ with 
states where $p$ is shifted by $\pm 1$, i.e. the states 
$e^{i(p+1)x}e^{-i\half\theta}$ and $e^{i(p-1)x}e^{i\frac{3}{2}\theta}$, whose 
unperturbed energies are $\half(p+1)^2 + \frac{1}{8\Lambda}$ and 
$\half(p-1)^2 + \frac{9}{8\Lambda}$, respectively. The potential has no 
diagonal matrix element, so the energy is unchanged to first order in $V_0$.

The eigenfunction of the Hamiltonian to first order in $V_0$, for the fixed 
value $p + \half$ of $r$, is
\begin{equation} 
\Psi(x,\theta) = e^{ipx}e^{i\half\theta} 
+ \frac{V_0}{2p+1} e^{i(p+1)x}e^{-i\half\theta} 
+ \frac{V_0}{-2p+1+\frac{2}{\Lambda}} e^{i(p-1)x}e^{i\frac{3}{2}\theta} \,,
\label{firstorderstate}
\end{equation}
where the denominators of the coefficients are proportional to 
differences between the energies of the unperturbed states.
The energy of the state $\Psi$, to second order in $V_0$ (found either by 
acting with the Hamiltonian, or by using the standard formula) is
\begin{equation}
E = \half p^2 + \frac{1}{8\Lambda} - \frac{V_0^2}{2(2p+1)} 
- \frac{V_0^2}{2(-2p+1+\frac{2}{\Lambda})} \,. 
\label{secorderenergy}
\end{equation}
This formula is valid, provided the unperturbed energy differences are 
not small compared to $V_0$. So $V_0$ must be much less than 1 and $p$ must not 
approach $-\half$. The perturbative 
approach therefore definitely fails for the states near $r=0$ that
we were considering earlier for fairly strong $V_0$. However, it is 
successful for small $p$, even if $\Lambda$ is not small and the last term 
of the formula (\ref{secorderenergy}) makes a significant contribution. 
Therefore, perturbation theory allows us to consider easily the 
spin $\frac{3}{2}$ contribution to low energy states, in contrast to 
our matrix method, which required this contribution to be negligible.

Let us compare our previous calculation of $\ep_-$, as a matrix
eigenvalue, with this perturbative estimate. From the expression
(\ref{evalues}), and converting it back to give the energy $E$ as 
a function of momentum $p$, we find, to second order in $V_0$, that
\begin{equation}
E = \half p^2 + \frac{1}{8\Lambda} - \frac{V_0^2}{2(2p+1)} \,,
\end{equation}
and this agrees with (\ref{secorderenergy}) provided $\Lambda$ is small. 
So the matrix method and perturbation theory agree where they should.

The conclusion is that perturbation theory is a good way to find
states of the disc in a certain regime, but that regime does not extend
to where spin-momentum coupling has the correlation we are seeking. In
the following sections we shall investigate the quantised three-dimensional
dynamics of a Skyrmion in a background potential. We should
expect the matrix method to be more effective than perturbation theory
for finding the desired form of spin-momentum coupling. We shall need
a model where the potential is fairly strong, and where states of the
Skyrmion with spin $\frac{3}{2}$ and higher are suppressed, relative
to the spin $\half$ states.

\section{Rolling on a half-filled lattice of Skyrmions}

Static solutions of the lightly bound Skyrme model are well-represented
by a point particle description \cite{ghkms17}, and we start by 
reviewing this. This approach is also expected to 
provide an accurate representation of dynamics, given that the same is true 
in a lower-dimensional toy model \cite{salmi&sutcliffe16}.  

A multi-Skyrmion modelling a nucleus with mass number $N$ 
is described by $N$ Skyrmion-like point particles, each with three
positional degrees of freedom and three orientational degrees of
freedom.  The rotational degrees of freedom could be expressed using an
SO(3) matrix, but for quantum mechanical calculations it is more
convenient to use an SU(2) matrix $q$.  Throughout this section we
will identify the group SU(2) with the group of unit quaternions,
making the identifications $\mathbf{i}=-\ii\sigma_1$,
$\mathbf{j}=-\ii\sigma_2$, $\mathbf{k}=-\ii\sigma_3$ between imaginary
quaternions and Pauli matrices.  The Lagrangian for the model consists
of standard kinetic terms for the positions and orientations, and
interaction potentials between pairs of particles (see \cite{ghkms17} for the
precise form).

The interaction potential is such that the particles tend to
arrange themselves into crystals with an FCC lattice structure, with a
preferred orientation at each lattice site.  In suitable length units
the FCC lattice is the set of vectors $(x,y,z)\in\ZZ^3$ such that
$x+y+z$ is even. The preferred orientation at lattice site
$(x,y,z)$ is $\mathbf{i}^x\mathbf{j}^y\mathbf{k}^z$.

We want to study the problem of a charge-1 Skyrmion rolling along the surface
of a half-filled FCC lattice.  See Figure 4. We assume that the lattice
sites with $x+y+z\leq-2$ are filled with particles in their preferred
orientations, and consider a Skyrmion moving freely in the plane
\begin{equation}
\Pi = \{ (x,y,z)\in\RR^3\::\: x+y+z = 0 \} \,.
\end{equation}
The degrees of freedom for this Skyrmion are its position coordinates
$(x,y,z)$ and its orientation $q\in\mathrm{SU}(2)$.  Its
dynamics can be described by a Lagrangian consisting of a standard
kinetic term and a potential function
$V:\Pi\times\mathrm{SU}(2)\to\RR$.  The kinetic terms are invariant
under the group $\mathrm{SU}(2)_I\times\mathrm{SU}(2)_S$ of
isorotations and rotations, with action
\begin{equation}
\label{spin isospin action}
\mathrm{SU}(2)_I\times\mathrm{SU}(2)_S \ni (g,h):(\vec{x},q)\mapsto (h\vec{x}h^{-1},gqh^{-1}) \,,
\end{equation}
where we identify vectors $\vec{x}\in\RR^3$ with imaginary quaternions
$x\mathbf{i}+y\mathbf{j}+z\mathbf{k}=-\ii
(x\sigma_1+y\sigma_2+z\sigma_3)$.  These terms are also invariant
under translations $(\vec{x},q)\mapsto(\vec{x}+\vec{c},q)$ and parity
transformations $(\vec{x},q)\mapsto(-\vec{x},q)$.

\begin{figure}
\begin{center}
\includegraphics[width=300pt]{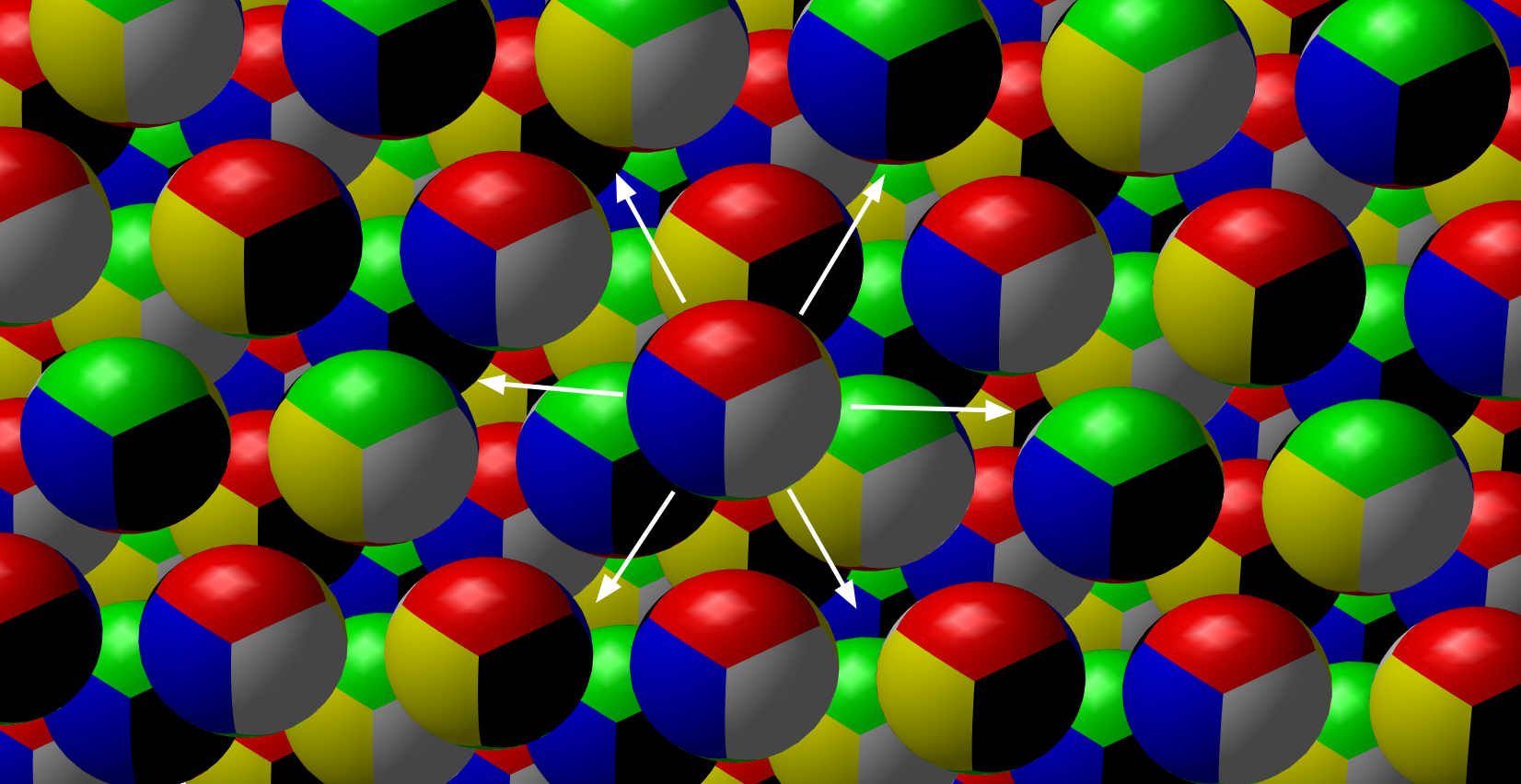}
\end{center}
\caption{A Skyrmion above a half-filled lattice of Skyrmions.  
The spheres are coloured using the colour scheme of \cite{ghs15} to indicate 
their orientations: the three pairs of opposite faces are coloured 
black/white, red/green and yellow/blue. Four orientations 
occur in the lattice. Preferred rolling directions for the single 
Skyrmion are shown by the arrows.}
\end{figure}

The potential function $V$ must be invariant under the group of
symmetries of the half-filled lattice.  This group is generated by the
following transformations:
\begin{align}
\label{sym1}-1:\big(x,y,z,q\big) &\mapsto \big(x,y,z,-q\big) \,, \\
\label{sym2}\rho:\big(x,y,z,q\big) &\mapsto
\big(z,x,y,\sfrac{1+\mathbf{i}+\mathbf{j}+\mathbf{k}}{2}
q\sfrac{1-\mathbf{i}-\mathbf{j}-\mathbf{k}}{2}\big) \,, \\
\label{sym3}\tau:\big(x,y,z,q\big) &\mapsto 
\big(x,y+1,z-1,\mathbf{i}q\big) \,, \\
\label{sym4}\sigma:\big(x,y,z,q\big) &\mapsto
\big(y,x,z,\sfrac{\mathbf{i}-\mathbf{j}}{\sqrt{2}}
q \sfrac{\mathbf{j}-\mathbf{i}}{\sqrt{2}} \big) \,.
\end{align}
The invariance of $V$ under $-\tau^2$, $-\rho\tau^2\rho^{-1}$ and
$-\rho^2\tau^2\rho^{-2}$ implies in particular that $V$ is invariant
under the translation action of the two-dimensional lattice
\begin{equation}
\Gamma=\{(2m,2n,-2(m+n))\::\:m,n\in\ZZ\}\subset\Pi \,.
\end{equation}
The transformations listed above acting on $(\Pi/\Gamma)\times
\mathrm{SU}(2)$ generate a finite group which is isomorphic to
the binary cubic group (the double cover of the cubic group).  Note
that since $\tau^2=-1$ when acting on
$(\Pi/\Gamma)\times\mathrm{SU}(2)$ we are free to use
$\rho,\sigma,\tau$ as a set of generators.  

We employ an ansatz for the potential of the form
\begin{equation}
\label{potential ansatz}
V(\vec{x},q) = U(\vec{x}) + \Tr(R(q)Y(\vec{x})) \,,
\end{equation}
with $R(q)$ the rotation matrix induced by $q$ (i.e.\
$q\sigma_jq^{-1}=\sigma_iR(q)_{ij}$), and $Y(\vec{x})$ a $3 \times 3$ 
matrix-valued function. This ansatz is motivated by the dipole description of
Skyrmion interactions; to a good approximation a single Skyrmion
interacts with a background field of pions like a triple of orthogonal
scalar dipoles, and this dipole interaction has similar $q$-dependence
to our ansatz.  Alternatively, one may regard our ansatz as the first
two terms in an expansion of $V$ in harmonics on
$\mathrm{SU}(2)$.  Note that this potential satisfies
$V(\vec{x},-q)=V(\vec{x},q)$, as required by symmetry.

We simplify the ansatz further using Fourier series. Both $U$ and $Y$
are required to be invariant under the lattice $\Gamma$, so have
Fourier series with summands corresponding to dual lattice vectors.
We assume that these Fourier series only contain terms corresponding
to the shortest dual lattice vectors; the associated functions are $1$
and $e^{\pm\ii\vec{a}_j.\vec{x}}$, where
\begin{equation}
\vec{a}_1 = \sfrac{\pi}{3}(2,-1,-1) \,, \ \vec{a}_2 =
\sfrac{\pi}{3}(-1,2,-1) \,, \ \vec{a}_3 = \sfrac{\pi}{3}(-1,-1,2) \,.
\end{equation}
With this restriction, the vector space of functions $U$ is
seven-dimensional and the space of functions $Y$ is 63-dimensional.

The symmetries $\rho$, $\sigma$, $\tau$ generate an action of the
binary cubic group on the vector spaces occupied by $U,Y$.  Since the ansatz 
\eqref{potential ansatz} is invariant under $q\mapsto-q$, this action descends 
to an action of the cubic group.
Representation theory can be used to find all functions $U$ and $Y$
which are invariant under this action.  This calculation involves the
irreducible representations of the cubic group: we recall these
briefly.  Besides the trivial representation $A_1$, there is another
one-dimensional representation $A_2$ in which $\rho$ and $\tau$ map to
1 and $\sigma$ maps to $-1$.  There is a unique two-dimensional
representation $E$ and two three-dimensional representations $T_1$ and
$T_2$; the first of these is the standard rotational action as the 
symmetry group of the cube and the second is $T_2=T_1\otimes A_2$.

It can be shown that the functions $e^{\pm\ii\vec{a}_j.\vec{x}}$
transform in the representation $2T_2$ of the cubic group; since this
contains no trivial subrepresentations the only allowed form for $U$
is a constant function.  Since this constant does not alter
differences between energy eigenvalues we set it to zero.

The elements of the group act on matrix-valued functions $Y$ by
simultaneously multiplying with matrices from the left and right, and
by permuting the Fourier modes.  The matrix acting from the left
corresponds to the representation $A_2\oplus E$, and that acting from
the right corresponds to the representation $T_1$.  The action on
Fourier modes is $A_1\oplus 2T_2$.  Therefore the representation
acting on the vector space occupied by $Y$ is $(A_1\oplus 2T_2)\otimes
T_1\otimes(A_2\oplus E)$.  Since $T_1\otimes (A_2\oplus E)\cong
T_1\oplus 2T_2$ this space contains four copies of $A_1$, so the space
of allowed potential functions has real dimension four.

This space of allowed potential functions can be parametrised by
$(U_0,U_1)=(W_0e^{\ii\theta_0},W_1e^{\ii\theta_1})\in\CC^2$ as
follows:
\begin{align}
Y_{\mathrm{approx}}(\vec{x}) &= -\Re \begin{pmatrix} U_0e^{\ii\vec{a}_1.\vec{x}} & U_1e^{\ii\vec{a}_2.\vec{x}} & U_1e^{\ii\vec{a}_3.\vec{x}} \\
U_1e^{\ii\vec{a}_1.\vec{x}} & U_0e^{\ii\vec{a}_2.\vec{x}} & U_1e^{\ii\vec{a}_3.\vec{x}} \\
U_1e^{\ii\vec{a}_1.\vec{x}} & U_1e^{\ii\vec{a}_2.\vec{x}} & U_0e^{\ii\vec{a}_3.\vec{x}}
\end{pmatrix} \nonumber \\
&=-\begin{pmatrix}
W_0\cos(\vec{a}_1.\vec{x}+\theta_0) & W_1\cos(\vec{a}_2.\vec{x}+\theta_1) & W_1\cos(\vec{a}_3.\vec{x}+\theta_1) \\
W_1\cos(\vec{a}_1.\vec{x}+\theta_1) & W_0\cos(\vec{a}_2.\vec{x}+\theta_0) & W_1\cos(\vec{a}_3.\vec{x}+\theta_1) \\
W_1\cos(\vec{a}_1.\vec{x}+\theta_1) &
W_1\cos(\vec{a}_2.\vec{x}+\theta_1) &
W_0\cos(\vec{a}_3.\vec{x}+\theta_0)
\end{pmatrix} \,.
\label{Y}
\end{align}
The values of the constants can be estimated in the lightly bound
Skyrme model using its point particle approximation.  One
calculates a function $Y_{\mathrm{true}}$ by adding up the interaction energies
between fixed Skyrmions in the planar lattice $x+y+z=-2$ and the
Skyrmion moving freely in the plane $x+y+z=0$, and then calculates its
Fourier coefficients.  The values obtained are
\begin{equation}
\label{potential parameters}
W_0 = 0.67 \,, \ W_1 = 0.55 \,, \ \theta_0 = -0.03 \,, \ \theta_1=0.77\approx
\frac{\pi}{4} \,.
\end{equation}
With these parameters the truncated Fourier series 
$Y_{\mathrm{approx}}$ given in eq.\ \eqref{Y} is a good approximation 
to $Y_{\mathrm{true}}$, in the sense that the ratio of the squares of
the $L^2$ norms of $Y_{\mathrm{true}}-Y_{\mathrm{approx}}$ 
and $Y_{\mathrm{true}}$ is 0.095. Our final potential $V(\vec{x},q) =
\Tr(R(q)Y_{\mathrm{approx}}(\vec{x}))$ is not exact, even
in the point particle description of Skyrmion interactions, but it is
analogous to the potential $-V_0\cos(x - \theta)$ that we chose for the
disc in section 2.

We claim that for the parameter values \eqref{potential parameters}, 
the potential given by equations \eqref{potential ansatz} and 
\eqref{Y} induces classical motion similar to a ball rolling on a surface.  
Consider the situation where a particle moves from 
$(x,y,z,q)=(0,0,0,1)$ to $(x,y,z,q)=(1,-1,0,\pm\mathbf{k})$.  
Both of these points are critical points of the potential, and for 
our parameter set they are minima. We will treat this situation 
adiabatically, assuming that the mass $M$ of the Skyrmion is 
much greater than its moment of inertia $\Lambda$.  
If the spatial kinetic energy $\sfrac12 Mv^2$ is much larger than 
the energy scale $W=\sqrt{W_0^2+W_1^2}$ of the potential 
then the path in space will to a good approximation be a 
straight line: $\vec{x}(t) = (vt/\sqrt{2})(1,-1,0)$.  
If the velocity is not too large then, at each time $t$, 
$q(t)$ will to a good approximation be the orientation 
that minimises $V(\vec{x}(t),q)$ with respect to variations in $q$.  
In this situation the rotational kinetic energy is roughly 
$\frac12 \Lambda v^2$, and the approximation is reliable as long as 
this is much less than $W$. Thus our approximation assumes that 
$W/M\ll v^2 \ll W/\Lambda$.

\begin{figure}
\begin{center}
\includegraphics[width=300pt]{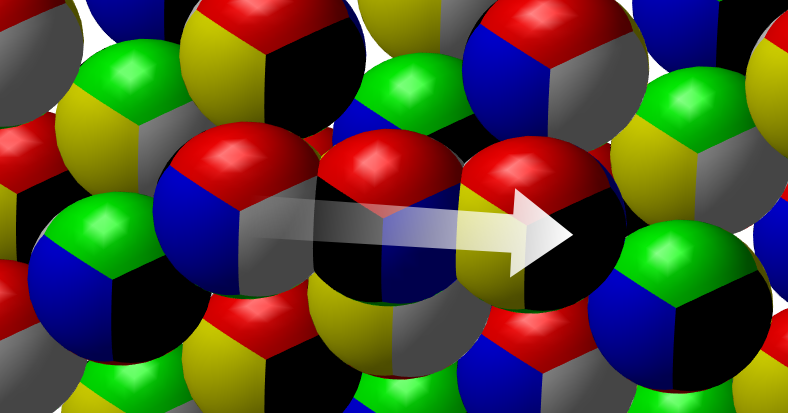}
\end{center}
\caption{The path of a rolling Skyrmion.}
\end{figure}

We wish to compare this motion with that of a rolling ball.  
If a ball of radius $r$ rolls with velocity $\vec{v}$ along a surface
with inward-pointing unit normal $\vec{n}$ its angular velocity 
will be $\vec{\omega}=-\vec{n}\times\vec{v}/r$. For 
$\vec{n}=(-1,-1,-1)/\sqrt{3}$ and $\vec{v}=(v/\sqrt{2})(1,-1,0)$ as 
above this makes the angular velocity a positive multiple of $(1,1,-2)$.  
The angle $\theta(t)$ between the angular velocity vector
$\vec{\omega}(t)=-2q^{-1}\dot{q}$ for the path $q(t)$ and the vector
$(1,1,-2)$ measures deviation from rolling motion: acute angles 
indicate motion similar to rolling, and obtuse angles indicate motion
that is opposed to rolling. We have computed $q(t)$ using the adiabatic 
approximation described above and have hence determined $\theta(t)$.  
The maximum angle along the path is $0.89\approx 2\pi/7$, indicating
that the motion induced by the potential is similar to that of a
rolling ball.

This adiabatic motion of a Skyrmion is illustrated in Figure 5 (see also Figure 4).  
The orientation of the rolling Skyrmion is illustrated at the start, mid-point, and end of the path.  
The start and end points are neighbouring lattice sites, and their orientations differ by a rotation of 180 degrees about the red-green axis.  
The most natural guess for the orientation at the mid-point is a rotation through 90 degrees about the same axis, 
and there are two possibilities here (depending on whether one rotates clockwise or anticlockwise).  
Figure 5 shows the orientation for one sense of rotation, but the alternative would have made the Skyrmion's red, white
and yellow faces visible at the mid-point. Now observe that just below the
Skyrmion at the mid-point there is a nearby Skyrmion in the lattice (white
and yellow faces visible). It is straightforward to find the pion dipole fields of this
pair of Skyrmions along the line joining them and verify that for the illustrated sense
of rotation, the fields are identical at the closest points, implying that the potential
energy is minimal. (The associated colouring is predominantly green, but with a
small tilt towards white and yellow.) If the sense of rotation had been opposite, the
field match would have been less good and the energy greater.

We conclude that the rolling motion illustrated in Figure 5 is along a particularly
deep valley in the potential energy landscape, and favoured as a low energy classical
motion. Anti-rolling is disfavoured.  Figure 5 suggests that to a good approximation the spin vector $\vec{S}$ for the 
rolling Skyrmion points in the direction of the red-green axis, from green to red.  This spin vector $\vec{S}$, the
vector $\vec{N}$ pointing into the half-filled lattice, and the momentum vector $\vec{P}$ do not form
an orthonormal triad ($\vec{P}$ is orthogonal to both $\vec{N}$ and $\vec{S}$ , and $\vec{N}.\vec{S} = -1/\sqrt{3}$), but
their triple scalar product $\vec{S}.\vec{N}\times\vec{P}$ is negative. This is what is expected classically
if the parameter $a$ in equation \eqref{coupling term} is positive.

\section{Weak coupling to the potential}

In this section and the next we will study the quantum mechanical problem of a
Skyrmion interacting with the surface of a half-filled lattice.  Since the
potential experienced by the Skyrmion is periodic it is natural to
analyse this problem using the theory of Bloch waves.  Let
$\vec{k}\in\RR^3$ be a crystal wavevector satisfying $k_1+k_2+k_3=0$
and let $\mathcal{H}_{\vec{k}}$ be the Hilbert space of wavefunctions
$\Psi:\RR^3\times\mathrm{SU}(2)\to\CC$ satisfying
\begin{align}
\Psi(\vec{x}+(t,t,t),q) &= \Psi(\vec{x},q) \quad\forall t\in\RR \,,
\\
\Psi(\vec{x}+\vec{v},q) &= e^{\ii\vec{k}.\vec{v}}\Psi(\vec{x},q)\quad\forall
\vec{v}\in\Gamma \,.
\end{align}
The first condition ensures that $\Psi$ is effectively defined in
the plane $\Pi$ rather than all of $\RR^3$.  The second condition has
the implication that two crystal wavevectors whose difference lies in
the reciprocal lattice $\Gamma^*$ generated by $\vec{a}_j$ define the
same Hilbert space, so $\vec{k}$ should be regarded as an element of
$\Pi^*/\Gamma^*$.

The natural operators on $\mathcal{H}_{\vec{k}}$ are isospin, spin,
and momentum.  Spin and isospin are just the infinitesimal versions of
the actions described in \eqref{spin isospin action}:
\begin{align}
S^j\Psi(\vec{x},q) &= \ii \frac{\dd}{\dd
  t}\Psi(\vec{x},q\exp(-\ii\sigma_jt/2))\bigg|_{t=0},\; \\
I^j\Psi(\vec{x},q) &= \ii \frac{\dd}{\dd
  t}\Psi(\vec{x},\exp(\ii\sigma_jt/2)q)\bigg|_{t=0} \,. 
\end{align}
Although the space in which the Skyrmion moves is two-dimensional, it
will be convenient to write momentum as a three-vector (due to the
three-dimensional origin of the problem).  Thus we set
\begin{equation} P^j\Psi(\vec{x},q) = -\ii\frac{\pa\Psi}{\pa x^j}(\vec{x},q) \,, \end{equation}
noting that $P^1+P^2+P^3=0$.  Then for a plane wave of the form
$\Psi(\vec{x},q)=e^{\ii\vec{b}.\vec{x}}$ with $b^1+b^2+b^3=0$ we
have $P^j\Psi=b^j\Psi$.

It will be useful in what follows to decompose $\mathcal{H}_{\vec{k}}$ 
into eigenspaces of $|\vec{S}|^2$.  Fix a non-negative integer or 
half-integer $\ell$  and let $\eta^\ell:\mathrm{SU}(2)\to
\mathrm{SU}(2\ell+1)$ be the spin $\ell$ irreducible representation of
$\mathrm{SU}(2)$.  If $\psi:\Pi\to\mathrm{Mat}(2\ell+1,\CC)$ is a
matrix-valued function of $\vec{x}$ then
\begin{equation} \Psi(\vec{x},q):=\Tr(\psi(\vec{x})\eta^\ell(q)) \end{equation}
satisfies $|\vec{S}|^2\Psi = |\vec{I}|^2\Psi = \ell(\ell+1)\Psi$.
Thus this wavefunction describes a particle of total spin $\ell$ and
total isospin $\ell$.  The space of all such wavefunctions in
$\mathcal{H}_{\vec{k}}$ will be denoted $\mathcal{H}_{\vec{k}}^\ell$.
The Peter-Weyl theorem implies that any wavefunction in
$\mathcal{H}_{\vec{k}}$ can be decomposed as an infinite sum of
wavefunctions of this type:
\begin{equation} \mathcal{H}_{\vec{k}} =
\bigoplus_{\ell\in\{0\}\cup\frac12\mathbb{N}}
\mathcal{H}_{\vec{k}}^\ell \,. \end{equation}

The wavefunction $\Psi$ describing the Skyrmion is required to
satisfy the Finkelstein--Rubinstein constraints \cite{anw83}.  These simply state
that $\Psi$ is an odd function of $q$:
$\Psi(\vec{x},-q)=-\Psi(\vec{x},q)$.  Functions in
$\mathcal{H}_{\vec{k}}^\ell$ are odd if $\ell$ is a half-integer and
even if $\ell$ is an integer.  Thus the Finkelstein--Rubinstein
constraints require $\Psi$ to be in the subspace
$\mathcal{H}_{\vec{k}}^{\rm odd}$ of $\mathcal{H}_{\vec{k}}$, where
the summation over $\ell$ is restricted to half-integers. This ensures
the quantised Skyrmion has half-integer spin.

\subsection{Outline of perturbation theory}

The hamiltonian that we will study is 
\begin{equation} H = H_0+V,\quad H_0 = \frac{|\vec{P}|^2}{2M} +
\frac{|\vec{S}|^2}{2\Lambda} \,. \end{equation}
Here $M,\Lambda>0$ are parameters representing the mass and moment of
inertia of the Skyrmion, and $V$ is the potential introduced in the
previous section.  We will construct an effective hamiltonian for the
lowest-energy eigenstates using perturbation theory, with the
parameters $W_0$ and $W_1$ of $V$ treated as small.

If $V=0$ and $\vec{k}$ is in the first Brillouin zone (i.e.\
$|\vec{k}+\vec{v}|> |\vec{k}|$ for all $\vec{v}\in\Gamma^\ast$) then
the lowest energy eigenstates in the space $\mathcal{H}_{\vec{k}}$ are
clearly of the form
\begin{equation}
\label{Psi0}
\Psi_0(\vec{x},q) = \Tr(\psi q)e^{\ii\vec{k}.\vec{x}} \,,
\end{equation}
with $\psi\in\mathrm{Mat}(2,\CC)$.  The space of all such
wavefunctions has dimension four and will be denoted by
$\mathcal{K}_{\vec{k}}$.  The energy of these states is
\begin{equation} E_0 = \frac{|\vec{k}|^2}{2M}+\frac{3}{8\Lambda} \,. \end{equation}
This is minimised by $\vec{k}=\vec{0}$.  In the following calculations
we will assume that $\vec{k}$ is close to $\vec{0}$, discarding terms
of $O(\vec{k}^2)$.

When the potential $V$ is non-zero, the four degenerate energy levels with 
energy $E_0$ will separate. We will study this effect using perturbation 
theory. Let us review the overall methodology, which generalises the
formulae (\ref{firstorderstate}) and (\ref{secorderenergy}). We seek 
an operator ${\cal I}:\mathcal{K}_{\vec{k}}\to\mathcal{H}_{\vec{k}}$ 
which depends continuously on the parameters $U_0,U_1$ in the potential, 
such that the image under $\cal I$ of the $H_0$-invariant subspace
$\mathcal{K}_{\vec{k}}$ is $H$-invariant, and such that the composition 
$\Pi_{\mathcal{K}}\,{\cal I}$ of $\cal I$ with the orthogonal projection
$\Pi_{\mathcal{K}}:\mathcal{H}_{\vec{k}}\to\mathcal{K}_{\vec{k}}$ is
the identity map.  The effective hamiltonian is then defined to be
$H_{\text{eff}}=\Pi_{\mathcal{K}}H{\cal I}$.  The operators ${\cal I}$ and
$H_{\text{eff}}$ will be constructed as power series in the parameters
that appear in the potential.

To zeroth order, $\cal I$ is just the inclusion:
${\cal I}|\Psi_0\rangle=|\Psi_0\rangle+O(V)$ for all
$\Psi_0\in\mathcal{H}_{\vec{k}}$.  The first order correction to
$H_{\text{eff}}$ is given by
 \begin{equation} H_{\text{eff}}|\Psi_0\rangle = \Pi_{\mathcal{K}}
 H|\Psi_0\rangle+O(V^2) = E_0|\Psi_0\rangle + \Pi_{\mathcal{K}}
 V|\Psi_0\rangle+O(V^2) \,. \end{equation}
The term linear in $V$ vanishes.  The reason for this is simple: the
only non-zero terms in the Fourier series of $V\Psi_0$ correspond to
plane waves of the form $e^{\ii(\vec{k}\pm\vec{a}_j).\vec{x}}$, as one
sees from eqs. \eqref{Psi0} and \eqref{Y}, and
these are all $L^2$-orthogonal to $e^{\ii\vec{k}.\vec{x}}$.  As a
consequence, $H_{\text{eff}}=H_0+O(V^2)$.

The first order correction to $\cal I$ is given by
\begin{equation} {\cal I}|\Psi_0\rangle = (1 - (H_0-E_0)^{-1}V)|\Psi_0\rangle+O(V^2). \end{equation}
This satisfies $H{\cal I}|\Psi_0\rangle = E_0{\cal I}|\Psi_0\rangle+O(V^2)$, so
its image is $H$-invariant up to terms quadratic in $V$.

The second order correction to $H_{\text{eff}}$ is given by
\begin{align}
 H_{\text{eff}}|\Psi_0\rangle &= \Pi_{\mathcal{K}}H(1 -
(H_0-E_0)^{-1}V)|\Psi_0\rangle+O(V^3) \nonumber
\\
 &= (E_0-\Pi_{\mathcal{K}}V(H_0-E_0)^{-1}V)|\Psi_0\rangle+O(V^3) \,.
\label{Heff2O}
\end{align}
In the next subsection we will calculate the action of $\Pi_{\mathcal{K}}V(H_0-E_0)^{-1}V$ on wavefunctions $\Psi_0$ of the
form \eqref{Psi0}, and thereby evaluate $H_{\text{eff}}$ to second order.  A reader uninterested in the details of this calculation may skip to the final result, eq.\ \eqref{finalHeff}.

\subsection{The effective hamiltonian $H_{\text{eff}}$}

We begin by analysing $V|\Psi_0\rangle$, with the potential $V$ given
by eqs. \eqref{potential ansatz} and \eqref{Y}. From 
eq. \eqref{potential ansatz} we see that
$V\in\mathcal{H}^1_{\vec{0}}$, and from eq.\ \eqref{Psi0} we see that
$\Psi_0\in\mathcal{H}_{\vec{k}}^{\frac12}$.  It follows from the
Clebsch--Gordan rules that the excited wavefunction $V(\vec{x},q)\Psi_0(\vec{x},q)$ will be a
sum of terms with spin $\frac12$ and spin $\frac32$.  Thus
\begin{equation} V|\Psi_0\rangle = \Pi^{\frac{1}{2}}V|\Psi_0\rangle +
\Pi^{\frac{3}{2}}V|\Psi_0\rangle \,, \end{equation}
with $\Pi^\ell$ denoting projection onto $\mathcal{H}_{\vec{k}}^\ell$.
Applying $(H_0-E_0)^{-1}$ to the spin $\frac12$ term gives
\begin{align}
(H_0-E_0)^{-1}\Pi^{\frac{1}{2}}V|\Psi_0\rangle
&= 2M(|\vec{P}|^2-|\vec{k}|^2)^{-1}\Pi^{\frac{1}{2}}V|\Psi_0\rangle
  \nonumber \\
&= 2M(|\vec{P}-\vec{k}|^2+2(\vec{P}
-\vec{k}).\vec{k})^{-1}\Pi^{\frac{1}{2}}V|\Psi_0\rangle \nonumber \\
&= 2M(|\vec{P}-\vec{k}|^2)^{-1} \Pi^{\frac{1}{2}}V|\Psi_0\rangle \nonumber\\
&\quad-4M(|\vec{P}-\vec{k}|^2)^{-2}\vec{k}.(\vec{P}-\vec{k})
  \Pi^{\frac{1}{2}}V|\Psi_0\rangle + O(\vec{k}^2) \,.
\end{align}
As the Fourier modes that appear in the excited wavefunction 
$V(\vec{x},q)\Psi_0(\vec{x},q)$ are
$e^{\ii(\vec{k}\pm\vec{a}_j).\vec{x}}$, the operator $|\vec{P}-\vec{k}|^2$
takes the constant value $|\vec{a}_j|^2=2\pi^2/3$ on
$\Pi^{\frac12}V|\Psi_0\rangle$, which simplifies this expression.  The spin
$\frac32$ term can be analysed in the same way, yielding
\begin{multline}
\label{splitting}
(H_0-E_0)^{-1}V|\Psi_0\rangle 
= \frac{3M}{\pi^2}\Pi^{\frac{1}{2}}V|\Psi_0\rangle 
- \frac{9M}{\pi^4}\vec{k}.(\vec{P}-\vec{k}) \Pi^{\frac{1}{2}}V|\Psi_0\rangle \\
+ M\left(\frac{\pi^2}{3}+\frac{3M}{2\Lambda}\right)^{-1}\Pi^{\frac{3}{2}}V|\Psi_0\rangle
- M\left(\frac{\pi^2}{3}+\frac{3M}{2\Lambda}\right)^{-2}\vec{k}.(\vec{P}-\vec{k})
\Pi^{\frac{3}{2}}V|\Psi_0\rangle \\
 + O(\vec{k}^2) \,.
\end{multline}
Thus to compute $\Pi_{\mathcal{K}}V(H_0-E_0)^{-1}V|\Psi_0\rangle$ we
need to compute the following four terms:
$\Pi_{\mathcal{K}}V\Pi^{\frac{1}{2}}V|\Psi_0\rangle$,
$\Pi_{\mathcal{K}}V\vec{k}.(\vec{P}-\vec{k})\Pi^{\frac{1}{2}}V|\Psi_0\rangle$,
$\Pi_{\mathcal{K}}V\Pi^{\frac{3}{2}}V|\Psi_0\rangle$ and
$\Pi_{\mathcal{K}}V\vec{k}.(\vec{P}-\vec{k})\Pi^{\frac{3}{2}}V|\Psi_0\rangle$.

We begin with $\Pi_{\mathcal{K}}V\Pi^{\frac{1}{2}}V|\Psi_0\rangle$.
This can be evaluated with the help of the following identity, which is proved in the appendix:
\begin{equation}
\label{Pi12}
\Pi^{\frac12}\big(R_{ji}(q)\Tr(\psi(\vec{x}) q)\big) =
\frac{1}{3}\Tr(\sigma_i\psi(\vec{x})\sigma_jq) \,.
\end{equation}
We introduce a vector
\begin{equation}
\vec{u}=(u_1,u_2,u_3)=(U_1,U_1,U_0)=(W_1e^{\ii\theta_1} \,,
  W_1e^{\ii\theta_1} \,, W_0e^{\ii\theta_0}) \end{equation}
so that
\begin{equation} V(\vec{x},q) = -\frac12\sum_{i,j=1}^3 (u_{i-j}e^{\ii\vec{a}_j.\vec{x}}
+\bar{u}_{i-j}e^{-\ii\vec{a}_j.\vec{x}})R_{ji}(q) \,, \end{equation}
with the index $i-j$ understood modulo 3.  Then applying the identity
\eqref{Pi12} yields
\begin{multline}
\Pi^{\frac12}\big(V\Psi_0)(\vec{x},q) = \\ 
 -\frac16 \sum_{i,j=1}^3 \left(u_{i-j}
  \Tr(\sigma_i\psi\sigma_jq)e^{\ii(\vec{k}+\vec{a}_j).\vec{x}} +
  \bar{u}_{i-j}
  \Tr(\sigma_i\psi\sigma_jq)e^{\ii(\vec{k}-\vec{a}_j).\vec{x}}
  \right) \,.
\end{multline}
To apply the operator $\Pi_{\mathcal{K}}V$ to this expression we
multiply the function with $V$, discard all terms in the Fourier
series except $e^{\ii\vec{k}.\vec{x}}$, and apply $\Pi^{\frac12}$ with
the help of the identity \eqref{Pi12}.  The result is
\begin{align}
\Pi_{\mathcal{K}}(V\Pi^{\frac12}\big(V\Psi_0))(\vec{x},q) &=
\frac{1}{36} \sum_{i,j,k=1}^3\left(\bar{u}_{k-j}u_{i-j} + u_{k-j}\bar{u}_{i-j} \right)
\Tr(\sigma_k\sigma_i\psi\sigma_j\sigma_jq)e^{\ii\vec{k}.\vec{x}} \nonumber \\
&= \frac{1}{36} \sum_{i,j,k=1}^3 \bar{u}_{k-j}u_{i-j}
\Tr((\sigma_k\sigma_i+\sigma_i\sigma_k)\psi q)e^{\ii\vec{k}.\vec{x}} \nonumber \\
&= \frac{1}{6} \sum_{i=1}^3|u_i|^2 \Tr(\psi q)e^{\ii\vec{k}.\vec{x}} \nonumber \\
&= \left(\frac{W_0^2}{6}+\frac{W_1^2}{3}\right)\Psi_0(\vec{x},q) \,. 
\label{id1}
\end{align}

The next term,
$\Pi_{\mathcal{K}}V\vec{k}.(\vec{P}-\vec{k})\Pi^{\frac{1}{2}}V|\Psi_0\rangle$,
can be evaluated using a similar method.  The calculation will make
use of the identity
\begin{equation} \vec{\bar{u}}\times \vec{u} 
= 2\sqrt{3}\ii\sin(\theta_1-\theta_0)\vec{n}\times\vec{e}_3 \,, \end{equation}
in which $\vec{e}_j$ are the standard basis vectors for $\RR^3$ and
\begin{equation} \vec{n} = -\frac{1}{\sqrt{3}}(1,1,1) \end{equation}
is an inward-pointing normal vector of unit length
representing the normalised gradient of the nuclear charge density.  Since
$\vec{k}.(\vec{P}-\vec{k})e^{\ii(\vec{k}\pm\vec{a}_j).\vec{x}} =
\pm\vec{k}.\vec{a}_je^{\ii(\vec{k}\pm\vec{a}_j).\vec{x}}$, we obtain
\begin{align}
\Pi_{\mathcal{K}}(V\vec{k}.(\vec{P}-\vec{k})&\Pi^{\frac12}\big(V\Psi_0))(\vec{x},q) \nonumber \\
&= \frac{1}{36} \sum_{i,j,k=1}^3 \vec{k}.\vec{a}_j\left(\bar{u}_{k-j}u_{i-j}  
- u_{k-j}\bar{u}_{i-j} \right)
\Tr(\sigma_k\sigma_i\psi\sigma_j\sigma_jq)e^{\ii\vec{k}.\vec{x}} \nonumber \\
&= \frac{1}{36} \sum_{i,j,k=1}^3 \vec{k}.\vec{a}_j\,\bar{u}_{k-j}u_{i-j}
\Tr((\sigma_k\sigma_i-\sigma_i\sigma_k)\psi q)e^{\ii\vec{k}.\vec{x}}
  \nonumber \\
&= -\frac{\sqrt{3}}{9}W_0W_1\sin(\theta_1-\theta_0) \sum_{j,l=1}^3
  \vec{k}.\vec{a}_j\,(\vec{n}\times\vec{e}_3)^{l-j}\Tr(\sigma_l\psi
  q)e^{\ii\vec{k}.\vec{x}} \,. 
\end{align}
Now $\sum_j\vec{k}.\vec{a}_j\,(\vec{n}\times\vec{e}_3)^{l-j}$
simplifies algebraically to $\pi(\vec{n}\times\vec{k})^l$ and
$\Tr(\sigma_l\psi q)e^{\ii\vec{k}.\vec{x}}=2 S^l\Psi_0$, so
\begin{equation}
\label{id2}
\Pi_{\mathcal{K}}V\vec{k}.(\vec{P}-\vec{k})\Pi^{\frac12}V|\Psi_0\rangle =
-\frac{2\pi}{3\sqrt{3}}W_0W_1\sin(\theta_1-\theta_0)
\vec{S}.\vec{n}\times\vec{k}|\Psi_0\rangle \,.
\end{equation}

The remaining two terms will be evaluated indirectly, using the identities
\begin{align}
\Pi_{\mathcal{K}}V\Pi^{\frac{3}{2}}V|\Psi_0\rangle 
&= \Pi_{\mathcal{K}}V^2|\Psi_0\rangle 
- \Pi_{\mathcal{K}}V\Pi^{\frac{1}{2}}V|\Psi_0\rangle \,, \\
\Pi_{\mathcal{K}}V\vec{k}.(\vec{P}-\vec{k})\Pi^{\frac{3}{2}}V|\Psi_0\rangle
&= \Pi_{\mathcal{K}}V\vec{k}.(\vec{P}-\vec{k})V|\Psi_0\rangle
- \Pi_{\mathcal{K}}V\vec{k}.(\vec{P}-\vec{k})\Pi^{\frac{1}{2}}V|\Psi_0\rangle \,.
\end{align}
In other words, we calculate the contributions from the sum of the
spin $\frac12$ and spin $\frac32$ excited states and subtract the spin
$\frac12$ contribution.

We begin with $\Pi_{\mathcal{K}}V^2|\Psi_0\rangle$.  The term in the
Fourier series of $V(\vec{x},q)^2\Psi_0(\vec{x},q)$ involving $e^{\ii\vec{k}.\vec{x}}$ is
\begin{equation}
\frac12\sum_{j=1}^3\left|\sum_{i=1}^3R_{ij}(q)u_{i-j}\right|^2\Tr(\psi
q)e^{\ii\vec{k}.\vec{x}} \,. \end{equation}
The other terms in the Fourier series will be annihilated by
$\Pi_{\mathcal{K}}$, so need not be computed.

By the Clebsch--Gordan rules, $|\sum_iR_{ij}(q)u_{i-j}|^2$ belongs to
the space
$\mathcal{H}_{\vec{0}}^0\oplus\mathcal{H}_{\vec{0}}^1\oplus\mathcal{H}_{\vec{0}}^2$
(because $R_{ij}(q)\in\mathcal{H}^1_{\vec{0}}$).  We only need to
calculate the piece in
$\mathcal{H}_{\vec{0}}^0\oplus\mathcal{H}_{\vec{0}}^1$, because
multiplying a spin $\frac12$ wavefunction with a spin $2$ function
yields wavefunctions with spin $\frac32$ and $\frac52$, both of which
will be annihilated by $\Pi_{\mathcal{K}}$.  We show in the appendix that, for any
vectors $\vec{v},\vec{w}\in\RR^3$,
\begin{equation} \Pi^0\big((v^iR_{ij}w^j)^2\big) =
\frac{1}{3}|\vec{v}|^2|\vec{w}|^2 \,,\quad
\Pi^1\big((v^iR_{ij}w^j)^2\big) = 0 \,. \end{equation}
Therefore the relevant part of $|\sum_iR_{ij}(q)u_{i-j}|^2$ is
$|\Re\vec{u}|^2+|\Im\vec{u}|^2=W_0^2+2W_1^2$.  It follows that
\begin{equation}
\Pi_{\mathcal{K}}V^2|\Psi_0\rangle =
\left(\frac{W_0^2}{2}+W_1^2\right)|\Psi_0\rangle
\end{equation}
and, using our earlier result \eqref{id1},
\begin{equation}
\label{id3}
\Pi_{\mathcal{K}}V\Pi^{\frac{3}{2}}V|\Psi_0\rangle =
\left(\frac{W_0^2}{3}+\frac{2W_1^2}{3}\right)|\Psi_0\rangle \,.
\end{equation}

The term $\Pi_{\mathcal{K}}V\vec{k}.(\vec{P}-\vec{k})V|\Psi_0\rangle$
can be evaluated using similar techniques.  The coefficient of
$e^{\ii\vec{k}.\vec{x}}$ in the Fourier series of
$(V\vec{k}.(\vec{P}-\vec{k})V\Psi_0)(\vec{x},q)$ is
\begin{equation}
\frac12 \sum_{j=1}^3 \vec{k}.\vec{a}_j\left|\sum_{i=1}^3
  R_{ij}(q)u_{i-j}\right|^2e^{\ii\vec{k}.\vec{x}}\Tr(\psi q) \,.
\end{equation}
As before, the other terms in the Fourier series are irrelevant.  Also
as before, we may replace $\left|\sum_i
  R_{ij}(q)u_{i-j}\right|^2$ with $W_0^2+2W_1^2$.  The resulting sum
over $j$ is zero, because $\sum_j\vec{a}_j=\vec{0}$.  Therefore
$\Pi_{\mathcal{K}}V\vec{k}.(\vec{P}-\vec{k})V|\Psi_0\rangle=0$ and, by
our previous result \eqref{id2},
\begin{equation}
\label{id4}
\Pi_{\mathcal{K}}V\vec{k}.(\vec{P}-\vec{k})\Pi^{\frac{3}{2}}V|\Psi_0\rangle
= \frac{2\pi}{3\sqrt{3}}W_0W_1\sin(\theta_1-\theta_0)\,
\vec{S}.\vec{n}\times\vec{k}|\Psi_0\rangle \,.
\end{equation}

We are now in a position to evaluate the effective hamiltonian.
Collecting together the results \eqref{Heff2O}, \eqref{splitting},
\eqref{id1}, \eqref{id2}, \eqref{id3} and \eqref{id4} gives
\begin{multline}
\label{finalHeff}
H_{\text{eff}} = \frac{|\vec{k}|^2}{2M} + \frac{3}{8\Lambda} -
M\left(\left(\frac{\pi^2}{3}\right)^{-1}+2\left(\frac{\pi^2}{3}
+\frac{3M}{2\Lambda}\right)^{-1}\right)\left(\frac{W_0^2}{6}+\frac{W_1^2}{3}\right)
\\ -M\left(\left(\frac{\pi^2}{3}\right)^{-2}-\left(\frac{\pi^2}{3}
+\frac{3M}{2\Lambda}\right)^{-2}\right)\frac{2}{3\sqrt{3}}W_0W_1
\sin(\theta_1-\theta_0)\vec{S}.\vec{n}\times\vec{k}
\\ + O(V^3) + O(\vec{k}^2) \,.
\end{multline}
This hamiltonian, which is analogous to equation 
\eqref{secorderenergy} in the 2D model, 
contains the sought-after coupling between momentum
and spin \eqref{coupling term}.  Besides scalars, this is the only
term in the hamiltonian, and it is at first sight surprising that no
other terms occur.  The explanation lies in the symmetries of the lattice:
$\vec{S}.\vec{n}\times\vec{k}$ is the only term linear in $\vec{k}$
which is invariant under the action of the binary cubic group.

For the parameter set \eqref{potential parameters} the coefficient of
the term \eqref{coupling term} in $H_{\mathrm{eff}}$ is negative,
which is opposite to what would be expected based on the classical
rolling motion of Skyrmions.  This is not such a surprise, given what
we learnt from the toy model.  In the toy model, spin-momentum effects
consistent with the classical rolling motion of Skyrmions only occurred
for a relatively strong potential, and were inaccessible to
perturbation theory.  In the next section we investigate stronger
potentials.

\section{Strong coupling to the potential}

In the previous section we discussed the situation where the potential
is small; in this section we discuss the case where the potential is
slightly larger.  Recall that in the 2D toy model, if the potential was
strong the lowest energy Bloch wave had a non-zero crystal wave
vector (at $r=0$ so $k = \pm\half$).  We expect a similar effect in 
the 3D model.  We begin this section by looking for candidate crystal 
wave vectors for the ground state, using symmetry as a guide.

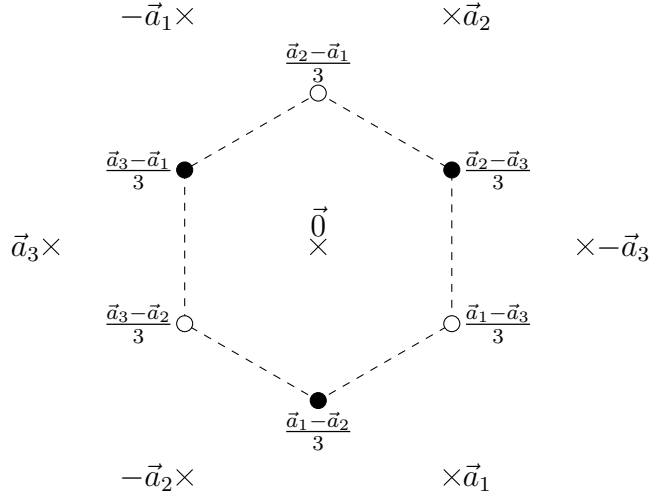
\begin{figure}[ht]
\begin{center}
\begin{tikzpicture}
\draw[dashed] (50pt,-29pt) node[right]{$\frac{\vec{a}_1-\vec{a}_3}{3}$} -- (50pt, 29pt) node[right]{$\frac{\vec{a}_2-\vec{a}_3}{3}$} -- (0pt,58pt) node[above]{$\frac{\vec{a}_2-\vec{a}_1}{3}$} -- (-50pt, 29pt) node[left]{$\frac{\vec{a}_3-\vec{a}_1}{3}$} -- (-50pt,-29pt) node[left]{$\frac{\vec{a}_3-\vec{a}_2}{3}$} -- (0pt,-58pt) node[below]{$\frac{\vec{a}_1-\vec{a}_2}{3}$} -- cycle;
\filldraw[fill=white] (50pt,-29pt) circle[radius=3pt];
\filldraw[fill=white] (0pt,58pt) circle[radius=3pt];
\filldraw[fill=white] (-50pt,-29pt) circle[radius=3pt];
\filldraw[fill=black] (50pt,29pt) circle[radius=3pt];
\filldraw[fill=black] (-50pt,29pt) circle[radius=3pt];
\filldraw[fill=black] (0pt,-58pt) circle[radius=3pt];
\draw (0pt,0pt) node[above]{$\vec{0}$};
\draw[-] (-3pt,-3pt) -- (3pt,3pt);
\draw[-] (-3pt,3pt) -- (3pt,-3pt);
\draw (-100pt,0pt) node[left]{$\vec{a}_3$};
\draw[-] (-103pt,-3pt) -- (-97pt,3pt);
\draw[-] (-103pt,3pt) -- (-97pt,-3pt);
\draw (50pt,87pt) node[right]{$\vec{a}_2$};
\draw[-] (53pt,84pt) -- (47pt,90pt);
\draw[-] (53pt,90pt) -- (47pt,84pt);
\draw (50pt,-87pt) node[right]{$\vec{a}_1$};
\draw[-] (53pt,-84pt) -- (47pt,-90pt);
\draw[-] (53pt,-90pt) -- (47pt,-84pt);
\draw (100pt,0pt) node[right]{$-\vec{a}_3$};
\draw[-] (103pt,-3pt) -- (97pt,3pt);
\draw[-] (103pt,3pt) -- (97pt,-3pt);
\draw (-50pt,87pt) node[left]{$-\vec{a}_1$};
\draw[-] (-53pt,84pt) -- (-47pt,90pt);
\draw[-] (-53pt,90pt) -- (-47pt,84pt);
\draw (-50pt,-87pt) node[left]{$-\vec{a}_2$};
\draw[-] (-53pt,-84pt) -- (-47pt,-90pt);
\draw[-] (-53pt,-90pt) -- (-47pt,-84pt);
\end{tikzpicture}
\end{center}
\caption{Diagram showing important vectors in crystal momentum space.  The dashed line indicates the boundary of the first Brillouin zone.  Shaded circles represent $\vec{k}_+$ and unshaded circles represent $\vec{k}_-$.}
\label{fig:brillouin}
\end{figure}

Recall that the hamiltonian is invariant under an action of the binary
cubic group.  The action of this group on wavefunctions induces an
action on the space of crystal wavevectors $\vec{k}$.  The generator
$\tau$ acts trivially on $\vec{k}$, while the generators $\rho$ and
$\sigma$ act on $\vec{k}$ as multiplication by the matrices
\begin{equation}
\begin{pmatrix}0&0&1\\1&0&0\\0&1&0\end{pmatrix}\quad\text{and}\quad 
\begin{pmatrix}0&1&0\\1&0&0\\0&0&1\end{pmatrix}.
\end{equation}
The vectors
\begin{equation} \vec{k}_+ =
\frac{1}{3}(\vec{a}_2-\vec{a_3})\quad\text{and}\quad\vec{k}_- =
\frac{1}{3}(\vec{a}_3-\vec{a_2}) \end{equation}
are special because they represent fixed points of the action of the
subgroup generated by $\rho$ and $\tau$, namely the binary tetrahedral
group (bear in mind that $\vec{k}$ is only defined up to addition of
the reciprocal lattice vectors $\vec{a}_j$).  These two crystal
wavevectors are plausible candidates for the wavefunction of the
ground state at strong coupling.  Note that they are at the vertices of
the first Brillouin zone, as shown in Figure \ref{fig:brillouin}.

In order to analyse the Hilbert spaces corresponding to these crystal
wavevectors it is convenient to apply a rotation to the lattice and
the moving Skyrmion:
\begin{equation} \big(\vec{x},q\big) \mapsto \big(R(U)^{-1}\vec{x},qU\big) \,,
\end{equation}
where
\begin{align} 
U &= \frac{1}{\sqrt{2(\sqrt{3}+3)}}\begin{pmatrix} (\sqrt{3}+1)e^{-\frac{\ii\pi}{8}} & -\sqrt{2}e^{-\frac{\ii\pi}{8}} \\ \sqrt{2}e^{\frac{\ii\pi}{8}} & (\sqrt{3}+1)e^{\frac{\ii\pi}{8}} \end{pmatrix}, \\
R(U)^{-1} &= \frac{1}{\sqrt{6}}\begin{pmatrix} 1&1&-2\\-\sqrt{3}&\sqrt{3}&0\\\sqrt{2}&\sqrt{2}&\sqrt{2}\end{pmatrix}.
\end{align}
After rotation, the Skyrmion moves in the plane
$z=0$, and the half-filled lattice of Skyrmions is the region $z<0$.
The generators of the binary cubic group now act as follows:
\begin{align}
-1: \big(x,y,z,q\big) & \mapsto \big(x,y,z,-q\big) \,, \\
\rho: \big(x,y,z,q\big) &\mapsto 
\big( -\sfrac12 x +\sfrac{\sqrt{3}}{2}y, -\sfrac12 y
+\sfrac{\sqrt{3}}{2}x,z ,\sfrac{1+\mathbf{i}+\mathbf{j}+\mathbf{k}}{2}
q\sfrac{1-\mathbf{k}\sqrt{3}}{2}\big) \,, \\
\tau:\big(x,y,z,q\big) &\mapsto 
\big(x-\sqrt{\sfrac32},y+\sfrac{1}{\sqrt{2}},z,\mathbf{i}q\big) \,, \\
\sigma:\big(x,y,z,q\big) &\mapsto
\big(x,-y,z,\sfrac{\mathbf{i}-\mathbf{j}}{\sqrt{2}}q\mathbf{j} \big) \,.
\end{align}
After rotation the reciprocal lattice vectors are
\begin{equation} \vec{a}_1 = \pi\sqrt{\frac{2}{3}}\begin{pmatrix} \frac{1}{2}\\-\frac{\sqrt{3}}{2}\\0 \end{pmatrix},\quad \vec{a}_2 = \pi\sqrt{\frac{2}{3}}\begin{pmatrix} \frac{1}{2}\\\frac{\sqrt{3}}{2}\\0 \end{pmatrix},\quad \vec{a}_3 = \pi\sqrt{\frac{2}{3}}\begin{pmatrix}-1\\0\\0 \end{pmatrix}. \end{equation}

\subsection{Perturbation theory in $\vec{k}$}
We will be interested in eigenfunctions of the hamiltonian whose
crystal wavevector is close to $\vec{k}_\pm$.  It is enough to analyse
just wavevectors close to $\vec{k}_+$, as the transformation
$\tau$ swaps $\vec{k}_+$ and $\vec{k}_-$.  First we will
identify an orthonormal basis $|\Psi_{0a}\rangle\in\mathcal{H}_{\vec{k}_+}$
for the eigenspace of the hamiltonian with (degenerate) lowest energy 
eigenvalue $E_0$.  Then we will consider nearby wavevectors
$\vec{k}=\vec{k}_++\delta\vec{k}$.  Perturbing $\vec{k}$ in this way
is mathematically equivalent to perturbing the momentum operator:
\begin{equation} \vec{P} \mapsto \vec{P}_0 + \delta\vec{k} \,, \end{equation}
where $\vec{P}_0=-\ii\nabla_{\vec{x}}$ is the usual momentum operator
acting on $\mathcal{H}_{\vec{k}_+}$.  Thus nearby wavevectors can be
analysed using perturbation theory.  The perturbed hamiltonian is
\begin{equation} H = H_0 + \frac{1}{M}\delta\vec{k}.\vec{P}_0 
+ \frac{1}{2M}|\delta\vec{k}|^2 \,,\quad 
H_0 = \frac{1}{2M}|\vec{P}_0|^2 + \frac{1}{2\Lambda}|\vec{S}|^2 + V \,. \end{equation}

We will show below that
$\langle\Psi_{0a}|\vec{P}_0|\Psi_{0b}\rangle=0$ for reasons of
symmetry, so the effective hamiltonian acting on this eigenspace is
unchanged to linear order in $\vec{k}$.  Therefore the perturbed
wavefunctions
\begin{equation} |\Psi_a\rangle = |\Psi_{0a}\rangle -
(H_0-E_0)^{-1}\frac{1}{M}\delta\vec{k}.\vec{P}_0|\Psi_{0a}\rangle+O(\delta\vec{k}^2) \end{equation}
satisfy $H|\Psi_a\rangle = E_0|\Psi_{a}\rangle+O(\delta\vec{k}^2)$.
We then compute the matrix elements of the hamiltonian $H$ to
second order in $\delta\vec{k}$:
\begin{multline}
\label{expectation of H}
\langle \Psi_{0a}|H|\Psi_b \rangle = E_0\delta_{ab} 
+ \frac{|\delta\vec{k}|^2}{2M}\delta_{ab} \\ 
- \frac{1}{M^2}\big\langle\Psi_{0a}\big|\delta\vec{k}.\vec{P}_0(H_0-E_0)^{-1}
\delta\vec{k}.\vec{P}_0\big|\Psi_{0b}\big\rangle + O(\delta\vec{k}^3) \,.
\end{multline}
For large enough $M$ the second term on the right dominates the third
term, meaning that the lowest energy eigenvalue has a stable local
minimum at $\delta\vec{k}=\vec{0}$.  Below we will quantify how large
$M$ needs to be for this to happen.

The expectation value $\langle\vec{P}\rangle$ of 
$\vec{P}=(P^1,P^2)$ in the state
$|\Psi_0\rangle$ is, as we have already noted, zero.  Similarly, group
theoretical arguments will show that the expectation value
$\langle\vec{S}\rangle$ of $\vec{S}=(S^1,S^2,S^3)$ has vanishing planar
components (although the component perpendicular to the plane will be
non-vanishing).  For $\delta\vec{k}\neq\vec{0}$ we expect these
expectation values to be non-zero and correlated.  More precisely, we expect 
$\langle\vec{P}_0\rangle$ to point in the same direction as $\vec{n}\times
\langle\vec{S}\rangle$, where $\vec{n}=(0,0,-1)$ is now the normalised 
gradient of the nuclear matter density.  Equivalently,
\begin{equation} \label{goal} \langle S^+\rangle = \lambda \ii \langle P_0^+\rangle\text{ for some }\lambda>0 \,, \end{equation}
where $S^\pm:=S^1\pm \ii S^2$ and $P_0^\pm:=P_0^1\pm\ii P_0^2$.

It is straightforward to derive expressions for these expectation
values within the framework of perturbation theory in $\delta\vec{k}$.  
The expectation value of $\vec{P}$ in a normalised state $v^a|\Psi_a\rangle$ is
$\bar{v}^av^b\langle\Psi_a|\vec{P}|\Psi_b\rangle$, where
\begin{multline}
\label{expectation of P}
\big\langle\Psi_a\big|\vec{P}\big|\Psi_b\big\rangle 
= \big\langle\Psi_a\big| (\vec{P}_0+\delta\vec{k})\big|\Psi_b\big\rangle \\
= \delta\vec{k}\,\delta_{ab} 
- \frac{1}{M}\big\langle\Psi_{0a}\big|\big(\vec{P}_0 (H_0-E_0)^{-1} 
\delta\vec{k}.\vec{P}_0 + \delta\vec{k}.\vec{P}_0 (H_0-E_0)^{-1} 
\vec{P}_0 \big) \big|\Psi_{0b}\big\rangle + O(\delta\vec{k}^2) \,.
\end{multline}
We will show below that for sufficiently large $M$ the second term is
negligible and we have that $\langle\vec{P}\rangle\approx\delta\vec{k}$.  For
$\vec{S}$ we compute
\begin{multline}
\big\langle\Psi_a\big| \vec{S}\big|\Psi_b\big\rangle \\
= - \big\langle\Psi_{0a}\big|\big(\vec{S} (H_0-E_0)^{-1} 
\delta\vec{k}.\vec{P}_0 + \delta\vec{k}.\vec{P}_0 (H_0-E_0)^{-1} 
\vec{S} \big) \big|\Psi_{0b}\big\rangle + O(\delta\vec{k}^2) \,.
\end{multline}
This equation and \eqref{expectation of P} are analogues 
of eqs. \eqref{lineariseExpectations} in the 2D model.  
In terms of $\kappa=\delta k_1+\ii\delta k_2$, we have that
$\langle P^+\rangle\approx\kappa$ and
\begin{multline}
\langle S^+\rangle = - \frac{\kappa}{2}\bar{v}^av^b\big\langle\Psi_{0a}\big|\big(S^+ (H_0-E_0)^{-1} P_0^- + P_0^- (H_0-E_0)^{-1} S^+ \big) \big|\Psi_{0b}\big\rangle \\
- \frac{\bar{\kappa}}{2}\bar{v}^av^b\big\langle\Psi_{0a}\big|\big(S^+ (H_0-E_0)^{-1} P_0^+ + P_0^+ (H_0-E_0)^{-1} S^+ \big) \big|\Psi_{0b}\big\rangle
\end{multline}
to leading order.  Thus to verify \eqref{goal} it is sufficient to show that
\begin{align}
\label{claim1}
\big\langle\Psi_{0a}\big|\big(S^+ (H_0-E_0)^{-1} P_0^- 
+ P_0^- (H_0-E_0)^{-1} S^+ \big) \big|\Psi_{0b}\big\rangle 
&= -\ii\lambda\delta_{ab},\;\lambda\in\RR_{>0} \,,\\
\label{claim2}
\big\langle\Psi_{0a}\big|\big(S^+ (H_0-E_0)^{-1} P_0^+ 
+ P_0^+ (H_0-E_0)^{-1} S^+ \big) \big|\Psi_{0b}\big\rangle &= 0 \,.
\end{align}
This concludes the outline of what we intend to show. In the remainder
of this section we verify equations \eqref{claim1} and \eqref{claim2}
by explicit calculation. In the next section we provide an alternative
verification based mainly on symmetry.

\subsection{Truncation of Hilbert space}
In order to calculate the eigenstates $|\Psi_{0a}\rangle$ we make a number of simplifying
assumptions.  First, we assume that the only terms that occur in the
spatial Fourier series of $\Psi_{0a}$ are those with the shortest
possible wavevectors, namely
\begin{equation} 
e_1(\vec{x}) := e^{\frac{\ii}{3}(\vec{a}_2-\vec{a}_3).\vec{x}} \,,\quad 
e_2(\vec{x}):=e^{\frac{\ii}{3}(\vec{a}_3-\vec{a}_1).\vec{x}} \,,\quad 
e_3(\vec{x}):=e^{\frac{\ii}{3}(\vec{a}_1-\vec{a}_2).\vec{x}} \,. 
\end{equation}
Note that these all have the same crystal wavevector; for example, in
the case of $e_1$ and $e_2$ this is because
\begin{equation} \frac{\vec{a}_2-\vec{a}_3}{3} - \frac{\vec{a}_3-\vec{a}_1}{3} 
= \frac{\vec{a}_1+\vec{a}_2+\vec{a}_3}{3}-\vec{a_3} = -\vec{a}_3 \,. \end{equation}

Second, we assume that the only terms that occur in the expansions of
$\Psi_{0a}$ in harmonics on SU(2) are those corresponding to spin $\half$.
In other words,
\begin{equation}
\Psi_{0a}(\vec{x},q) = \Tr(\psi_{a}(\vec{x})q) = \Tr( \psi_{ia}e_i(\vec{x})q)
\end{equation}
for $2\times 2$ matrices $\psi_{1a},\psi_{2a},\psi_{3a}$.  Since these three matrices 
have altogether 12 degrees of freedom, the eigenstates  
$|\Psi_{0a}\rangle$ belong to a 12-dimensional subspace of the
Hilbert space $\mathcal{H}_{\vec{k}_+}$.

These assumptions are justified as long as energies of states in
the 12-dimensional subspace are appreciably lower than those in its
complement.  If the moment of inertia $\Lambda$ is small then states with
spin greater than $\half$ will have much greater energy than the spin $\half$
states considered here, so truncation to spin $\half$ can always be
justified by choosing $\Lambda$ small.  To justify the truncation in
momentum space, we need to consider the next-shortest wavevectors
associated with $\vec{k}_+$.  These are
$\frac{2}{3}(\vec{a}_3-\vec{a}_2)$, $\frac{2}{3}(\vec{a}_1-\vec{a}_3)$
and $\frac{2}{3}(\vec{a}_2-\vec{a}_1)$, and their associated kinetic
energies are
\begin{equation}
\label{next energy level} 
\frac{1}{2M}\left\|\frac{2}{3}(\vec{a}_2-\vec{a}_1)\right\|^2+\frac{3}{8\Lambda} 
= \frac{4\pi^2}{9M}+\frac{3}{8\Lambda} \,. 
\end{equation}
Later we will compare these with the energies of states in the 12-dimensional subspace.

The generators $r=\rho,\tau$ of the binary tetrahedral group act
naturally on wavefunctions $\mathcal{H}_{\vec{k}_+}$ via
$r\cdot\Psi(\vec{x},q)=\Psi(r^{-1}(\vec{x}, q))$, and these actions
fix the 12-dimensional subspace.  However, they only define a
projective representation and not a true representation, because
\begin{multline}
\tau^2\cdot \Psi(\vec{x},q) = \Psi\big(x+\sqrt{6},y-\sqrt{2},z,-q\big) = e^{\ii(\sqrt{6},-\sqrt{2},0).\vec{k}_+}\Psi(\vec{x},-q) \\ 
= e^{2\pi\ii/3} \Psi(\vec{x},-q)\neq\Psi(\vec{x},-q) \,.
\end{multline}
The binary tetrahedral group is known to be Schur-trivial, meaning
that every projective representation can be turned into a true
representation by twisting the actions of the group elements.  
In this case, a true representation is obtained by choosing
\begin{equation} \rho\cdot\Psi(\vec{x},q) = \Psi(\rho^{-1}(\vec{x}, q)) \,, \quad 
\tau\cdot\Psi(\vec{x},q) = \omega \Psi(\tau^{-1}(\vec{x}, q)) \,. \end{equation}
Here we have introduced $\omega=e^{2\pi\ii/3} =
-\frac{1}{2}+\ii\frac{\sqrt{3}}{2}$, the cube root of unity.

We wish to break up the 12-dimensional subspace of the Hilbert space
into irreducible subrepresentations of the binary tetrahedral
group.  To this end, we review these irreducible representations.  
Besides the trivial representation, there are two further
1-dimensional representations $A_a$ with $a=1,2$, given by
\begin{equation} -1\mapsto 1 \,,\quad \rho\mapsto \omega^a \,,\quad \tau\mapsto 1 \,. \end{equation}
The binary tetrahedral group can be identified with the subgroup of
the group of unit quaternions generated by $-1$,
$\rho=-\frac12(1+\mathbf{i}+\mathbf{j}+\mathbf{k})$,
$\tau=\mathbf{i}$.  The standard identification of unit quaternions
with SU(2) matrices gives a two-dimensional representation $E_3$.  There
are two further inequivalent representations $E_1=E_3\otimes A_1$ and
$E_2=E_3\otimes A_2$.  Finally, there is a three-dimensional
representation $F$ given by $R:\mathrm{SU}(2)\to\mathrm{SO}(3)$.

It is straightforward to check that the action of the binary
tetrahedral group on the span of
$e_1,e_2,e_3\in\mathcal{H}_{\vec{k}_+}$ is isomorphic to the
representation $F$.  The action on the four-dimensional subspace of
$\mathcal{H}_{\vec{0}}$ consisting of functions of the form
$\Psi(\vec{x},q)=\Tr(\psi q)$ is isomorphic to $E_1\oplus E_2$.  This
can be seen as follows: the induced action on the $2\times 2$ matrix
$\psi$ is
\begin{align}
\rho\cdot\psi &= \sfrac{-1-\mathbf{k}\sqrt{3}}{2}\psi\left(\sfrac{-1-\mathbf{i}-\mathbf{j}-\mathbf{k}}{2}\right)^{-1}\\
&= \begin{pmatrix}\omega & 0 \\ 0 & \omega^2\end{pmatrix}\psi
\left(\frac{-1+\ii\sigma_1+\ii\sigma_2+\ii\sigma_3}{2}\right)^{-1} \,, \\ 
\tau\cdot\psi &= \psi \mathbf{i}^{-1} = \psi(-\ii\sigma_1)^{-1} \,, \\
-1\cdot\psi &= \psi(-1)^{-1} \,.
\end{align}
The matrices acting on the left correspond to the representation
$A_1\oplus A_2$, and those acting on the right correspond to the
representation $E_3$, so the representation is $E_3\otimes(A_1\oplus
A_2)\cong E_1\oplus E_2$.

The action on our 12-dimensional subspace is therefore
$F\otimes(E_1\oplus E_2)$.  This, it turns out, is isomorphic to
$2E_3\oplus 2E_1\oplus 2E_2$.  To fully describe the decomposition, we
introduce basis vectors $f_{ia}$, with $i=1,\ldots 6$ and $a=1,2$:
\begin{align}
f_{1a}
  &= \begin{pmatrix}\delta_{1a}&\delta_{2a}\\0&0\end{pmatrix}
(-\omega\ii\sigma_1e_1-\omega^2\ii\sigma_2e_2-\ii\sigma_3e_3) \,, \\
f_{2a} &= \begin{pmatrix}0&0\\\delta_{1a}&\delta_{2a}\end{pmatrix}
(-\omega^2\ii\sigma_1e_1-\omega\ii\sigma_2e_2-\ii\sigma_3e_3) \,, \\
f_{3a} &= \begin{pmatrix}\delta_{1a}&\delta_{2a}\\0&0\end{pmatrix}
(-\ii\sigma_1e_1-\ii\sigma_2e_2-\ii\sigma_3e_3) \,, \\
f_{4a} &= \begin{pmatrix}0&0\\\delta_{1a}&\delta_{2a}\end{pmatrix}
(-\omega\ii\sigma_1e_1-\omega^2\ii\sigma_2e_2-\ii\sigma_3e_3) \,,
\end{align}
\begin{align}
f_{5a} &= \begin{pmatrix}\delta_{1a}&\delta_{2a}\\0&0\end{pmatrix}
(-\omega^2\ii\sigma_1e_1-\omega\ii\sigma_2e_2-\ii\sigma_3e_3) \,, \\
f_{6a} &= \begin{pmatrix}0&0\\\delta_{1a}&\delta_{2a}\end{pmatrix}
(-\ii\sigma_1e_1-\ii\sigma_2e_2-\ii\sigma_3e_3) \,.
\end{align}
It can be checked that $f_{1a}$ span an irreducible subrepresentation
isomorphic to $E_3$, $f_{2a}$ span a second copy of $E_3$, $f_{3a}$ and
$f_{4a}$ span two copies of $E_1$, and $f_{5a}$ and $f_{6a}$ span two
copies of $E_2$.

\subsection{Hamiltonian matrix and the ground state}
Next we need the matrix elements \eqref{expectation of H} for the hamiltonian 
acting on our truncated Hilbert space.  The non-trivial part is the
potential.  After rotation, the potential given by equations 
\eqref{potential ansatz} and \eqref{Y} becomes
\begin{equation} V(\vec{x},q) = -\Re\Tr\left(\frac{1}{\sqrt{3}}(U_0+2U_1)A_0(\vec{x})R(q) + \sqrt{\frac{2}{3}}(U_0-U_1)A_1(\vec{x})R(q) \right), \end{equation}
where $U_{\alpha}=W_{\alpha}e^{\ii\theta_{\alpha}}$ ($\alpha = 0,1$)
as before and
\begin{align}
A_0(\vec{x}) &= \begin{pmatrix}0&0&0\\0&0&0\\e^{\ii\vec{a}_1.\vec{x}}&e^{\ii\vec{a}_2.\vec{x}}&e^{\ii\vec{a}_3.\vec{x}}\end{pmatrix}, \\
A_1(\vec{x}) &= \begin{pmatrix}\frac{1}{2}e^{\ii\vec{a}_1.\vec{x}}&\frac{1}{2}e^{\ii\vec{a}_2.\vec{x}}&-e^{\ii\vec{a}_3.\vec{x}}\\ -\frac{\sqrt{3}}{2}e^{\ii\vec{a}_1.\vec{x}}&\frac{\sqrt{3}}{2}e^{\ii\vec{a}_2.\vec{x}}&0 \\ 0&0&0 \end{pmatrix}.
\end{align}
This acts on wavefunctions from our 12-dimensional space by
multiplication, and, in order to have a well-defined action, the 
resulting functions need to be projected back onto the 12-dimensional space.

Consider first the action of the functions $e^{\ii\vec{a}_j.\vec{x}}$
with $j=1,2,3$.  In the case $j=1$ we find that
\begin{align}
e^{\ii\vec{a}_1.\vec{x}}e_1 &= e^{\ii(3\vec{a}_1+\vec{a}_2-\vec{a}_3).\vec{x}/3} \\
e^{\ii\vec{a}_1.\vec{x}}e_2 &= e^{\ii(2\vec{a}_1+\vec{a}_3).\vec{x}/3}
=e^{\ii(\vec{a}_1-\vec{a}_2).\vec{x}/3}=e_3\\
e^{\ii\vec{a}_1.\vec{x}}e_3 &= e^{\ii(4\vec{a}_1-\vec{a}_2).\vec{x}/3} \,.
\end{align}
The first and third of these are orthogonal to $e_1,e_2,e_3$ so only
the second of these survives projection onto the span of
$e_1,e_2,e_3$.  By performing similar computations we find that the
actions of $e^{\ii\vec{a}_j.\vec{x}}$ are
\begin{equation}
\label{action of Fourier modes}
e^{\ii\vec{a}_j.\vec{x}} e_k = \delta_{j+1,k} \, e_{k+1} 
\quad\text{(no sum over $k$) \,.}
\end{equation}
In this expression, indices $i,j,k$ are to be understood modulo 3.

The effect of multiplying a wavefunction with $R_{ij}(q)$ and 
projecting back to the 12-dimensional space is described by the identity 
\eqref{Pi12}.  Therefore the action of the functions 
$\Tr(A_{\alpha}(\mathbf{x})R(q))$ that appear in
the potential on the 12-dimensional subspace of the Hilbert
space can be computed using equations \eqref{action of Fourier modes}
and \eqref{Pi12}, and turns out to be
\begin{equation} \Tr(A_{\alpha}(\mathbf{x})R(q)) \cdot \psi_{ib} =
B_{\alpha;ji}\psi_{jb} \,, \end{equation}
where $B_{\alpha}$ are $6\times 6$ block diagonal matrices of the form
\begin{equation} B_{\alpha} = \left(\begin{array}{c|c|c} C_{\alpha} & 0 & 0 \\ \hline 0 
& \omega C_{\alpha} & 0 \\ \hline 0 & 0 & \omega^2C_{\alpha} \end{array} \right),\quad 
C_0=\frac{1}{3}\begin{pmatrix} -\ii\omega^2 & 0 \\ 0 & \ii\omega \end{pmatrix},\quad C_1 = \frac{1}{3}\begin{pmatrix} 0&\ii\\\ii&0\end{pmatrix}.
\end{equation}
The action of the potential function is therefore described by the $6\times6$
block diagonal matrix
\begin{equation} -\frac{1}{2}\left( \frac{1}{\sqrt{3}}(U_0+2U_1)B_0 + \sqrt{\frac{2}{3}}(U_0-U_1)B_1+ \text{hermitian conj.\ }\right). \end{equation}
It is straightforward to find the eigenvalues and eigenvectors for the
values of $U_0,U_1$ given earlier.  The lowest eigenvalue turns out to
be $-0.38$ and the associated eigenvectors are
\begin{equation} \psi_{0a} = -\mu f_{3a} + \nu f_{4a},\quad \mu = 0.46,\quad
\nu = 0.89,\quad a=1,2 \,. 
\label{groundstate-psi0a}
\end{equation}

Let us compare the energy of this state with the energy of the state
associated with $\vec{k}=0$.  The former is
\begin{equation} \frac{|\vec{k}_+|^2}{2M}+\frac{3}{8\Lambda} - 0.38 =
\frac{\pi^2}{9M}+\frac{3}{8\Lambda} - 0.38 \,. \end{equation}
The latter was computed in the previous section using perturbation theory to be
\begin{equation} \frac{3}{8\Lambda}- \frac{W_0^2+2W_1^2}{12}\left(\left(\frac{\pi^2}{3M}\right)^{-1}+\left(\frac{\pi^2}{3M}+\frac{3}{2\Lambda}\right)^{-1}\right). \end{equation}
The value of $W_0^2+2W_1^2$ is approximately 1.06.  Since we are only
interested in energy differences we ignore the term $3/8\Lambda$ which
occurs in both expressions.  Since we have been assuming
that $\Lambda$ is small, the other $\Lambda$-dependent term in brackets can be 
ignored.  Thus the state with crystal wavevector $\vec{k}_+$ will 
have lower energy if
\begin{equation} \frac{\pi^2}{9M} - 0.38 < -1.06\frac{M}{4\pi^2} \,. \end{equation}
This inequality holds for $M$ in the range $4.04<M<10.11$.  
Thus for $M$ close to zero (equivalent to small potentials) the state with
$\vec{k}=\vec{0}$ is preferred, but as $M$ increases past the value
$4.04$ the state with $\vec{k}=\vec{k}_+$ is preferred.

Now we assess the reliability of the approximation that we made by
truncating in momentum space.  The largest eigenvalue of the
$6\times6$ block diagonal matrix that describes the potential is 0.37.
Thus the largest energy involved in our calculation is
\begin{equation} \frac{|\vec{k}_+|^2}{2M}+\frac{3}{8\Lambda} + 0.37 
= \frac{\pi^2}{9M}+\frac{3}{8\Lambda} + 0.37 \,. \end{equation}
In our truncation of the Hilbert space we neglected states whose energy
is bounded below by \eqref{next energy level}.  
We are justified in neglecting these provided that
\begin{equation} \frac{\pi^2}{9M}+ 0.37 \ll \frac{4\pi^2}{9M} \iff M \ll
\frac{\pi^2}{3\times0.37} \approx 8.9 \,. \end{equation}
This means that our approximation is valid for the values of $M$
around 4.04 where the transition between the $\vec{k}=\vec{0}$ and 
$\vec{k}=\vec{k}_+$ states occurs.


\subsection{Expectation values for spin and momentum}
Now we turn our attention to the expectation value of spin and momentum in the
ground state.  We need to compute matrices describing the action of
$\vec{P}_0$ and $\vec{S}$ on the 12-dimensional subspace of the
Hilbert space.

The action of $\vec{S}$ is given by
\begin{equation} S^j \Tr(\psi(\vec{x})q) = \frac{1}{2}\Tr(\sigma_j\psi(\vec{x})q) \,. \end{equation}
It follows that $S^3 f_{ia} = (-1)^{i+1}\frac{1}{2}f_{ia}$ for
$i=1,\ldots,6$ and $a=1,2$.  In particular, for the ground states
$\psi_{0a}$ given by \eqref{groundstate-psi0a} we have 
$S^3\psi_{0a} = \frac{1}{2}(-\mu f_{3a}-\nu
f_{4a})$, and the expectation value of $S^3$ is
$\frac{1}{2}(\mu^2-\nu^2)<0$.  The action of $S^+=S^1+\ii S^2$ is
described by the $6\times 6$ matrix
\begin{equation} \left( \begin{array}{c|c|c} 0 & \frac{1}{2}\sigma_+ & 0 \\ \hline 0
& 0 & \frac{1}{2}\sigma_+ \\ \hline
\frac{1}{2}\sigma_+ & 0 & 0 \end{array} \right),\quad
\frac{1}{2}\sigma_+ = \frac{1}{2}(\sigma_1+\ii\sigma_2)
= \begin{pmatrix}0&1\\0&0\end{pmatrix}. \end{equation}
The action of $S^-$ is given by the conjugate transpose of this matrix. As 
the blocks on the diagonal are zero, the expectation values of $S^1$
and $S^2$ in the states $\psi_{0a}$ are zero, so the expected spin points
vertically down into the half-filled lattice of Skyrmions, as was previously claimed.

The action of $\vec{P}_0=-\ii\nabla$ on the functions $e_1$, $e_2$,
$e_3$ is simply
\begin{equation} \vec{P}_0e_1 = \frac{1}{3}(\vec{a}_2-\vec{a}_3),\quad 
\vec{P}_0e_2 = \frac{1}{3}(\vec{a}_3-\vec{a}_1),\quad 
\vec{P}_0e_3 = \frac{1}{3}(\vec{a}_1-\vec{a}_2) \,. \end{equation}
It follows that the action of $P_0^+=P_0^1+\ii P_0^2$ is described 
by the $6\times6$ block diagonal matrix
\begin{equation} -\ii\frac{\pi\sqrt{2}}{3} \left( \begin{array}{c|c|c} 
0 & I_2 & 0 \\ \hline 0 & 0 & I_2 \\ 
\hline I_2 & 0 & 0 \end{array} \right),\quad 
I_2=\begin{pmatrix}1&0\\0&1\end{pmatrix}. \end{equation}
The action of $P_0^-$ is given by the hermitian conjugate of this
matrix.  It follows that the expectation value of $\vec{P}$ in the
ground state is zero as claimed.

Using these formulae it is straightforward to verify equations
\eqref{claim1} and \eqref{claim2}.  We have that
\begin{align}
S^+\psi_{0a} &= \nu f_{1a} \,, & P_0^+ \psi_0^a 
&= -\ii\frac{\pi\sqrt{2}}{3}\left(-\mu f_{1a}+\nu f_{2a} \right)
  \,, \\
S^-\psi_{0a} &= -\mu f_{6a} \,, & P_0^- \psi_0^a 
&= -\ii\frac{\pi\sqrt{2}}{3}\left(-\mu f_{5a}+\nu f_{6a} \right) \,.
\end{align}
The block diagonal structure of the matrix representing $H_0$ means
that the inner products on the left hand side of \eqref{claim2} vanish as
required.  Using these identities and our particular values for
$U_0,U_1$, we find that
\begin{align}
\langle\Psi_{0a}|P_0^-(H_0-E_0)^{-1}S^+|\Psi_{0b}\rangle &=
\delta_{ab}\ii\frac{\pi\sqrt{2}}{3}
\begin{pmatrix}-\mu & \nu \end{pmatrix} \begin{pmatrix}
1.49 & 0.29 \\
0.29 & 1.89 \end{pmatrix} \begin{pmatrix}\nu \\ 0 \end{pmatrix}
\nonumber \\
&= -0.37\frac{\pi\sqrt{2}}{3}\ii\delta_{ab} \,, \\
\langle\Psi_{0a}|S^+(H_0-E_0)^{-1}P_0^-|\Psi_{0b}\rangle 
&= \delta_{ab}\ii\frac{\pi\sqrt{2}}{3}
\begin{pmatrix} 0 & -\mu \end{pmatrix} 
\begin{pmatrix} 4.88 & 0.09 \\ 0.09 & 1.93  \end{pmatrix} 
\begin{pmatrix}-\mu\\ \nu \end{pmatrix} \nonumber \\
&= -0.76\frac{\pi\sqrt{2}}{3}\ii\delta_{ab} \,.
\end{align}
Thus equation \eqref{claim1} holds true with $\lambda=1.13\pi\sqrt{2}/3$, 
a positive number.

It remains to evaluate the subleading contributions to $\langle\vec{P}\rangle$
and $E_0$.  The subleading term in \eqref{expectation of P} is
expressed in terms of
\begin{equation} T^{ij}_{ab} = \frac{1}{M}\big\langle
\Psi_{0a}\big|P_0^i(H_0-E_0)^{-1}P_0^j
+P_0^j(H_0-E_0)^{-1}P_0^i\big|\Psi_{0b}\big\rangle \,,\quad i,j=1,2 \,. \end{equation}
Note that by construction $T^{ij}_{ab}=T^{ji}_{ab}$.  It is
straightforward to show using the matrix given earlier for $P_0^+$ that
$\langle\Psi_{0a}|P_0^+(H_0-E_0)^{-1}P_0^+|\Psi_{0b}\rangle=0$ and
$\langle\Psi_{0a}|P_0^-(H_0-E_0)^{-1}P_0^-|\Psi_{0b}\rangle=0$.  These
two identities imply that $T^{11}_{ab}=T^{22}_{ab}$ and
$T^{12}_{ab}=-T^{21}_{ab}$.  Altogether, this means that $T^{ij}_{ab}$
is proportional to $\delta^{ij}$.  The coefficient can be determined
by evaluating
\begin{align}
2\big\langle \Psi_{0a}\big|&P_0^i(H_0-E_0)^{-1}P_0^i\big|\Psi_{0b}\big\rangle \nonumber \\
&= \big\langle
  \Psi_{0a}\big|P_0^+(H_0-E_0)^{-1}P_0^-\big|\Psi_{0b}\big\rangle 
+ \big\langle \Psi_{0a}\big|P_0^-(H_0-E_0)^{-1}
P_0^+\big|\Psi_{0b}\big\rangle \nonumber \\
&= \frac{2\pi^2}{9}\delta_{ab}
\begin{pmatrix}-\mu & \nu\end{pmatrix} 
\begin{pmatrix} 1.49 & 0.29 \\ 0.29 & 1.89 \end{pmatrix} 
\begin{pmatrix}-\mu \\ \nu \end{pmatrix} \nonumber \\
& \qquad +\frac{2\pi^2}{9}\delta_{ab} 
\begin{pmatrix} -\mu & \nu \end{pmatrix} 
\begin{pmatrix} 4.88 & 0.09 \\ 0.09 & 1.93 \end{pmatrix} 
\begin{pmatrix}-\mu\\ \nu \end{pmatrix} \nonumber \\
&= 4.03 \, \delta_{ab}.
\end{align}
Thus $T^{ij}_{ab}=4.03 \, \delta_{ab}\delta^{ij}$, and
\begin{equation} \big\langle \Psi_{0a}\big|\vec{P}\big|\Psi_{0b}\big\rangle 
= \left(1-\frac{4.03}{M}\right)\delta_{ab}\delta\vec{k} 
+ O(\delta\vec{k}^2) \,. \end{equation}
So for $M<4.03$, the expectation of momentum points in the opposite
direction to $\delta\vec{k}$ and for $M>4.03$, the region of most
interest, they point in the same direction. Notice that the 
transition occurs at almost exactly the same value of $M$ as where the 
energy for $\vec{k}=\vec{k}_+$ drops below that for $\vec{k}=\vec{0}$.

Finally we consider the subleading corrections to the eigenvalue $E$
of $H$ implied by eq.\ \eqref{expectation of H}.  This equation can be
rewritten in terms of $T^{ij}_{ab}$ as follows:
\begin{equation}
\frac{\langle\Psi_{a}|H|\Psi_b\rangle}{\langle\Psi_{a}|\Psi_b\rangle}
= E_0\delta_{ab}+\frac{\delta k_i\delta k_j}{M^2}
(M\delta^{ij}\delta_{ab} - T^{ij}_{ab}) \,.
\end{equation}
Inserting our formula for $T^{ij}_{ab}$ shows that $E=E_0 +
(M-4.03)\|\delta\vec{k}\|^2/M^2+O(\delta\vec{k}^3)$.  Thus
$\delta\vec{k}=0$ is a stable critical point when $M>4.03$.

This concludes our verification of spin-momentum coupling based on the
crystal wavevector $\vec{k}_+$.  If $M>4.04$ then the crystal wavevector $\vec{k}_+$ is 
preferred over $\vec{k}=\vec{0}$, and the expectation values
of spin and momentum are correlated in the manner predicted by the spin-momentum coupling.  
Our calculation is reliable as long as $M\ll8.9$.

\section{Symmetry arguments}

To conclude, we would like to point out that our results in the previous section
are robust and insensitive to the details of the choice of potential
function.  Many of them can be derived using symmetry alone, as
we now explain.

We begin by analysing the symmetry properties of the operators
$\vec{P}$ and $\vec{S}$.  Their commutation relations with $\rho$ are 
as follows:
\begin{equation} \rho S^3 = S^3\rho \,,\quad \rho S^+ = \omega^2S^+\rho \,,\quad 
\rho P^+=\omega^2 P^+\rho \,. \end{equation}
Since the hamiltonian commutes with the action of the binary
tetrahedral group, the eigenspace corresponding to the lowest
eigenvalue $E_0$ forms a representation $K$ of this group.
Generically this representation will be irreducible, as was the case
in the above calculation.  Since the group element $-1$ acts
non-trivially on the Hilbert space, $K$ must be isomorphic to one of
the three representations $E_3$, $E_1$ and $E_2$ introduced above, because
$-1$ acts trivially in all other irreducible representations of the
binary tetrahedral group.  The commutation relations above show that
the images of $K$ under $S^+$ and $P^+$ are isomorphic to $K\otimes
A_2$.  Since tensoring with $A_2$ cyclically permutes the representations $E_3$, $E_1$ and
$E_2$, these image representations are not isomorphic to $K$. It follows
that they are orthogonal to $K$. This means that
\begin{equation} \langle \Psi_{0a}|S^+|\Psi_{0b}\rangle =
0\quad\text{and}\quad\langle \Psi_{0a}|P_0^+|\Psi_{0b}\rangle = 0 \,, \end{equation}
and in particular that $S^+$ and $P_0^+$ have zero expectation value
in the ground state.

The identity \eqref{claim2} can be proved similarly.  The operators
$S^+(H_0-E_0)^{-1}P^+$ and $P^+(H_0-E_0)^{-1}S^+$ map $K$ onto a
representation isomorphic to $K\otimes A_1$, which is again not
isomorphic to $K$, so the inner products in \eqref{claim2} have to
vanish.

To analyse the identity \eqref{claim1} we need the symmetry
$\sigma$.  As has already been noted, $\sigma$ maps the Hilbert space
$\mathcal{H}_{\vec{k}_+}$ onto $\mathcal{H}_{\vec{k}_-}$.  There is
another transformation which swaps $\vec{k}_+$ and $\vec{k}_-$, namely
time reversal $T$.  This acts as
\begin{equation} T:\Psi(\vec{x},q)\mapsto \overline{\Psi}(\vec{x},q) \,. \end{equation}
The composition $\sigma T$ maps $\mathcal{H}_{\vec{k}_+}$ onto
$\mathcal{H}_{\vec{k}_+}$.  Its commutation relations with $\vec{S}$
and $\vec{P}$ are
\begin{equation} \sigma T P_0^1 = -P_0^1\sigma T \,,\quad
\sigma T P_0^2 = P_0^2\sigma T \,,\quad
\sigma T S^1 = S^1\sigma T \,,\quad
\sigma T S^2 = -S^2\sigma T \,. \end{equation}
Since multiplication with $\ii$ anticommutes with $\sigma T$, the
transformation $\sigma T$ anticommutes with $P^\pm$ and commutes with $S^\pm$.

The operator that appears in \eqref{claim1} is
$S^+(H_0-E_0)P_0^-+P_0^-(H_0-E_0)^{-1}S^+$.  When composed with
projection onto the eigenspace $K$ it defines a linear map $K\to K$.
This map commutes with the action of $\rho$ and $\tau$, so by Schur's
lemma it acts as multiplication by a scalar.  Since it anticommutes
with the action of $\sigma T$, this scalar must be pure imaginary.

Thus symmetry arguments show that an identity similar to
\eqref{claim1} must hold, with $\lambda\in\RR$.
However, symmetry arguments alone cannot determine the sign of $\lambda$.  This
is because replacing the potential $V$ with its negative $-V$ changes
the sign of $\lambda$ without altering the symmetry properties.
Nevertheless, the sign of $\lambda$ does seem to be fixed by a few coarse
features of the above calculation.  Consider again the basis vectors
$\psi_{0a}$ for the lowest-energy eigenspace.  Each of these can be
written as a sum of three terms:
\begin{multline}
\psi_{0a} = 
-\begin{pmatrix}-\mu\delta_{1a} & -\mu\delta_{2a} \\ 
\omega\nu\delta_{1a} & \omega\nu\delta_{2a} \end{pmatrix}\ii\sigma_1 e_1 \\
-\begin{pmatrix}-\mu\delta_{1a} & -\mu\delta_{2a} \\ 
\omega^2\nu\delta_{1a} & \omega^2\nu\delta_{2a} \end{pmatrix}\ii\sigma_2 e_2
-\begin{pmatrix}-\mu\delta_{1a} & -\mu\delta_{2a} \\
\nu\delta_{1a} & \nu\delta_{2a} \end{pmatrix}\ii\sigma_3 e_3 \,.
\end{multline}
Each summand is an eigenvector of $\vec{P}_0$, so has a definite
momentum vector.  Each summand also determines a unique spin vector
$\vec{v}$, such that it is an eigenstate of $\vec{v}.\vec{\sigma}$
acting from the left with eigenvalue $\half$.  The momentum vectors and
spin vectors for the summands involving $e_1$, $e_2$ and $e_3$ are
listed below:
\begin{equation*}
\begin{array}{ccc}
\text{summand} & \text{momentum vector} & \text{spin vector} \\
e_1 & \frac{\pi\sqrt{2}}{3}
\left(\frac{\sqrt{3}}{2},\,\frac{1}{2},\,0\right) 
& \left(\frac{1}{2}\mu\nu,\,-\frac{\sqrt{3}}{2}\mu\nu,\,
\frac{1}{2}(\mu^2-\nu^2)\right) \\
e_2 & \frac{\pi\sqrt{2}}{3}
\left(-\frac{\sqrt{3}}{2},\,\frac{1}{2},\,0\right) 
& \left(\frac{1}{2}\mu\nu,\,\frac{\sqrt{3}}{2}\mu\nu,\,
\frac{1}{2}(\mu^2-\nu^2)\right) \\
e_3 & \frac{\pi\sqrt{2}}{3}\left(0,\,-1,\,0\right) 
& \left(-\mu\nu,\,0,\,\frac{1}{2}(\mu^2-\nu^2)\right).
\end{array}
\end{equation*}
Note that for each summand, the momentum vector points in the opposite
direction to the cross product of $\vec{n}$ with the spin vector.

The expectation values for momentum and spin are weighted averages
of these vectors.  In the case $\delta \vec{k}=\vec{0}$ the three
summands contribute equally to the wavefunction, and weighted averages
are ordinary averages.  Since the momentum vectors sum to zero and the
unweighted average of the spin vectors is
$\frac{1}{2}(\nu^2-\mu^2)\vec{n}$, we recover the results derived
earlier.  When $\delta\vec{k}\neq\vec{0}$ the momentum eigenvalues get
shifted by $\delta\vec{k}$ and the dominant contribution to the
wavefunction is from the summand with the shortest wavevector.  For
example, when
$\delta\vec{k}$ points in the direction $\left(-\frac{\sqrt{3}}{2},\,-\frac{1}{2},\,0\right)$
the dominant contribution is from the state with momentum vector
aligned with $-\delta\vec{k}$, so the expectation value for
$(S^1,S^2)$ points in the direction
$\left(\frac{1}{2},\,-\frac{\sqrt{3}}{2}\right)$ and $\vec{n}\times
\langle\vec{S}\rangle$ points in the direction of $\delta\vec{k}$.  There are two
effects contributing to the expectation value for $\vec{P}$: the shift
in momentum vectors and the change of weights.  For strong potentials
the former dominates, and the expectation value for $\vec{P}$ points
in the same direction as the naive momentum $\delta\vec{k}$ (see the
discussion around eq.\ \eqref{expectation of P}).  Thus $\vec{n}\times
\langle\vec{S}\rangle$ and $\langle\vec{P}\rangle$ point in the same direction, consistent
with the spin-momentum coupling.

Note that all of this follows from the correlation between the spin
and momentum vectors of the three summands making up $\psi_{0a}$, and
any vector similar to $\psi_{0a}$ with $\mu\nu>0$ would produce
the same effect.  Thus we expect a similar correlation between spin
and momentum for all values of $U_0$, $U_1$ close to those used in our
calculation.

\section{Conclusions and further work}

In a classical picture, the experimentally observed nuclear spin-orbit
coupling arises from a rolling motion of a nucleon over the surface
of a larger nucleus. However, understanding why such a rolling motion
is energetically preferred remains something of a mystery. We have 
shown here that for a Skyrmion close to the planar surface of a 
half-filled lattice of Skyrmions, a rolling motion is energetically 
favoured by the orientational part of the potential energy. To
describe this planar rolling motion, it is convenient to introduce 
the notion of spin-momentum coupling.

We have next investigated the quantum mechanics of the
Skyrmion, first by analysing the hamiltonian describing the 
Skyrmion interacting with the half-filled lattice of Skyrmions 
using perturbation theory. A spin-momentum coupling 
term appears at second order in perturbation theory, but has the wrong 
sign, at least for the parameter set obtained from the lightly bound 
Skyrme model. We then calculated spin-momentum coupling at the 
level of expectation values, and found that the correct sign is 
recovered non-perturbatively at stronger potential strengths. The 
change of sign is correlated with a jump in the crystal momentum 
of the lowest energy state.

Our results were based on a half-filled FCC lattice that has been
sliced in the plane $x+y+z=0$.  There is another natural way to slice
the FCC lattice, in a plane parallel to one of the coordinate planes
(or $x=0$, $y=0$ or $z=0$).  It would be interesting to investigate
the spin-momentum coupling in that situation.

Our analysis also sheds light on a recent study of a $B=1$ Skyrmion orbiting
a $B=4$ core \cite{gudnason&halcrow18}.  It was found that weak
pion-induced coupling to the core affects the energy levels of the
orbiting Skyrmion, but in the opposite way to what would be expected
based on the phenomenological spin-orbit coupling. This is consistent with our
perturbative result for the spin-momentum coupling, and a similar 
problem will likely persist for larger baryon numbers. We suggest 
that the correct sign of the spin-orbit coupling will be obtained 
for stronger potentials, and that a non-perturbative treatment 
will resolve some of the puzzles in \cite{gudnason&halcrow18}.  

\medskip
\noindent\textbf{Acknowledgements}  This collaboration was initiated at the workshop Analysis of
Gauge-Theoretic Moduli Spaces at the Banff International Research Station, 
and we thank the organisers Rafe Mazzeo, Michael Singer and Sergey Cherkis. 
NSM thanks the School of Mathematics, University of Leeds for hospitality. 
NSM's work has been partially supported by STFC consolidated grant ST/P000681/1.

\appendix

\section{Identities for products of SU(2) harmonics}

In this appendix we prove two identities for products of harmonic
functions on $\mathrm{SU}(2)$.  To prove them, it is helpful to
identify $\mathrm{SU}(2)$ with $S^3\subset\RR^4$ by writing
\begin{equation} \mathrm{SU}(2)\ni q = q_0-\sum_{j=1}^3q_j\ii\sigma_j \sim
(q_0,q_1,q_2,q_3)\in S^3 \,. \end{equation}
If $p(q)$ is any homogeneous polynomial function on $\RR^4$ of degree
$2\ell$ that solves Laplace's equation then the restriction to $S^3$
lies in the space $\mathcal{H}^\ell$ of harmonic functions with total
spin and isospin $\ell$, because
\begin{equation} 0 = \triangle p = \frac{\pa^2 p}{\pa r^2} + \frac{3}{r}\frac{\pa
  p}{\pa r} - 4 |\vec{S}|^2p = 4\ell(\ell+1)p-4|\vec{S}|^2p \,. \end{equation}
So for example $q_0^2\notin \mathcal{H}^1$, because 
$\triangle q_0^2=2$, but $q_0^2-q_1^2\in\mathcal{H}^1$.

The first identity to be proved is
\begin{equation} \Pi^{\frac12}R_{ij}(q)\Tr(\psi q) 
= \frac13\Tr(\sigma_j\psi\sigma_i q) \,. \end{equation}
It is enough to prove this in the case $i=j=3$, as the other cases can
be deduced from this one by acting on $q$ with
$\mathrm{SU}(2)_I\times\mathrm{SU}(2)_S$.  From the definition
$q\sigma_jq^{-1}=\sigma_iR_{ij}(q)$ one deduces that $R_{33}(q) =
q_0^2-q_1^2-q_2^2+q_3^2$.  We calculate:
\begin{align}
q_0R_{33}(q) &= \frac{1}{3}q_0(q_0^2+q_1^2+q_2^2+q_3^2) + \frac23
q_0(q_0^2-2q_1^2-2q_2^2+q_3^2) \,, \\
q_1R_{33}(q) &= -\frac{1}{3}q_1(q_0^2+q_1^2+q_2^2+q_3^2) + \frac23
q_1(2q_0^2-q_1^2-q_2^2+2q_3^2) \,, \\
q_2R_{33}(q) &= -\frac{1}{3}q_2(q_0^2+q_1^2+q_2^2+q_3^2) + \frac23
q_2(2q_0^2-q_1^2-q_2^2+2q_3^2) \,, \\
q_3R_{33}(q) &= \frac{1}{3}q_3(q_0^2+q_1^2+q_2^2+q_3^2) 
+ \frac23 q_3(q_0^2-2q_1^2-2q_2^2+q_3^2) \,.
\end{align}
In each case the first term is in $\mathcal{H}^{\frac12}$ and the
second is in $\mathcal{H}^{\frac32}$.  Since
$\sigma_3q\sigma_3=q_0+q_1\ii\sigma_1+q_2\ii\sigma_2-q_3\ii\sigma_3$,
the result follows.

The second set of identities are
\begin{equation} \Pi^0((v^iR_{ij}(q)w^j)^2) = 
\frac13 |\vec{v}|^2|\vec{w}|^2 \,,\quad 
\Pi^1((v^iR_{ij}(q)w^j)^2)=0 \,, \end{equation}
for vectors $\vec{v},\vec{w}\in\RR^3$.  Again, by symmetry it is
enough to prove these in the case $\vec{v}=\vec{w}=(0,0,1)$.  We compute:
\begin{multline}
(R_{33}(q))^2 = \frac13(q_0^2+q_1^2+q_2^3+q_3^2)^2 \\ 
+ \frac23(q_0^4+q_1^4+q_2^4+q_3^4+2q_0^2q_3^2+2q_1^2q_2^2
-4(q_0^2+q_3^2)(q_1^2+q_2^2)) \,.
\end{multline}
The first term is $\frac{1}{3}$ as required and it is straightforward to check
that the second term solves Laplace's equation so belongs to
$\mathcal{H}^2$.

\end{document}